\documentclass[11pt]{article}
\usepackage{jheppub}

\usepackage{graphicx}

\usepackage{amsthm,amsmath,amssymb,bbm}
\usepackage{dsfont}  % \mathds{1}
\usepackage{enumitem} % \begin{itemize}[leftmargin=*,noitemsep]

\makeatletter
\def\@fpheader{\relax}
\makeatother

\preprint{IGC-18/12-1\\[-3em]}

\newlength{\bracewidth}

\usepackage{amsmath,amsfonts,amssymb,amsthm}
\usepackage{graphicx}
\usepackage{hyperref}
\usepackage{dsfont}
\usepackage{subcaption}
\usepackage{color}

\newcommand{\ket}[1]{\left|#1\right\rangle}
\newcommand{\bra}[1]{\left\langle #1\right|}

\newcommand{\Tr}{\mathrm{Tr}}

\newcommand{\BB}{\left|\partial A \right|}

\usepackage{float}
\usepackage[section]{placeins}

\usepackage{algorithm}
\usepackage[noend]{algpseudocode}

\makeatletter
\def\BState{\State\hskip-\ALG@thistlm}
\makeatother

\toccontinuoustrue

\title{\huge Entanglement entropy of Bell-network states in loop quantum gravity: Analytical and numerical results}

\author[a]{Eugenio Bianchi,}
\emailAdd{ebianchi@gravity.psu.edu}
\author[a]{\;Pietro Don\`a,}
\emailAdd{pxd81@psu.edu}
\author[b]{\;Ilya Vilensky}
\emailAdd{ilya.vilensky@fau.edu}

\affiliation[a]{Institute for Gravitation and the Cosmos \& Physics Department,\\ Penn State, University Park, PA 16802, USA}
\affiliation[b]{Florida Atlantic University, 777 Glades Road, Boca Raton, FL 33431, USA}

\abstract{Bell-network states are loop-quantum-gravity states that glue quantum polyhedra with entanglement. We present an algorithm and a code that evaluates the reduced density matrix of a Bell-network state and computes its entanglement entropy. In particular, we use our code for simple graphs to study properties of Bell-network states and to show that they are non-typical in the Hilbert space. Moreover, we investigate analytically Bell-network states on arbitrary finite graphs. We develop methods to compute the R\'enyi entropy of order p for a restriction of the state to an arbitrary region. In the uniform large-spin regime, we determine bounds on the entanglement entropy and show that it obeys an area law. Finally, we discuss the implications of our results for correlations of geometric observables.
}

\begin{document}
\maketitle
\flushbottom

\section{Introduction}
Elementary quanta of space---also known as quantum polyhedra \cite{Bianchi:2010gc}---are a characterizing feature of loop quantum gravity (LQG) \cite{Rovelli:2004tv,Ashtekar:2004eh,Thiemann:2007pyv,Rovelli:2014ssa}. These microscopic degrees of freedom are ultra-local and associated with the $N$ nodes of a finite graph $\Gamma$. The collective state of the system is an element of the kinematical Hilbert space of the theory which, at fixed spin-network graph $\Gamma$ and spins $j_\ell$, has the tensor-product structure
\begin{equation}
\label{LQGHspace}
\mathcal{H}_{\Gamma j_\ell}=\bigotimes_{n\in\Gamma}\mathcal{H}_n \, .
\end{equation}
Here $\mathcal{H}_n$ is the Hilbert space of the $SU(2)$ intertwiner associated to each node $n$, the quantization of a classical polyhedron \cite{Bianchi:2010gc,Barbieri:1997ks,Baez:1999tk,Bianchi:2011ub,Bianchi:2012wb}. This decomposition is the basis of the geometric picture of quantum space: an LQG state is a many-body state of quantum polyhedra with the adjacency relations given by the connectivity of the graph and areas given by the spins \cite{Rovelli:2010km}. A generic state in the space $\mathcal{H}_{\Gamma j_\ell}$ is a linear superposition of quantum polyhedra, 
\begin{equation}
|s\rangle=\sum_{i_1,\ldots, i_N}c_{i_1\ldots i_N}\;|i_1\rangle\cdots|i_N\rangle \, .
\label{eq:s-state}
\end{equation}
The factorized orthonormal basis associated to the tensor product (\ref{LQGHspace}), denoted $|\Gamma,j_\ell,i_n\rangle=|i_1\rangle\cdots|i_N\rangle$, is called the spin-network basis. This is a basis of simultaneous eigenstates of of a maximal commuting set of operators that are \emph{ultralocal}, i.e., each operator measures a geometric property of a single quantum polyhedron such as its volume or the dihedral angle between two faces \cite{Rovelli:1994ge,Ashtekar:1997fb,Barbieri:1997ks,Major:1999mc,Dittrich:2008va,Bianchi:2008es,Rovelli:2010km,Bianchi:2010gc,Baez:1999tk,Bianchi:2011ub,Bianchi:2012wb}. As a result, spin-network basis states are factorized over polyhedra: they are un-entangled. On the other hand a typical state of this many-body system has the form \eqref{eq:s-state} and represents entangled polyhedra.

\medskip

The connectivity of the graph $\Gamma$, together with the factorized structure of the Hilbert space $\mathcal{H}_{\Gamma j_\ell}$, allows us to define regions of the graph and their associated Hilbert space. Specifically, we call $A$ a region if the set of nodes $n\in A$  is path connected with respect to the graph structure. The associated Hilbert space is $\mathcal{H}_A=\bigotimes_{n\in A}\mathcal{H}_n$ and, denoting $\bar{A}$ the complement of $A$, we have a bipartition of the Hilbert space as the tensor product
\begin{equation}
\mathcal{H}_{\Gamma j_\ell}=\mathcal{H}_{A\phantom{\bar{A}}}\!\!\!\!\otimes \mathcal{H}_{\bar{A}}\,.
\end{equation}
Given this structure, it is immediate to define the entanglement entropy $S_A$ of a pure state $|s\rangle$ restricted to the subsystem $A$. Let us assume that $1\ll \dim \mathcal{H}_A\leq\sqrt{\dim \mathcal{H}_{\Gamma j_\ell}}$. In this case, Page's result \cite{PageTyp} states that a typical state in the Hilbert space $\mathcal{H}_{\Gamma j_\ell}$ has entanglement entropy 
\begin{equation}
S_A(\textrm{typical})\;\approx\; \log (\dim \mathcal{H}_A)\;-\;\frac{1}{2}(\dim \mathcal{H}_A)^2/\dim \mathcal{H}_{\Gamma,j_\ell}\,.
\label{eq:Page}
\end{equation}
This result indicates that the restriction $\rho_A$ of a typical pure state in $\mathcal{H}_{\Gamma j_\ell}$ is close to being maximally mixed when the subsystem $A$ is small. It is instructive to  consider the case of a graph $\Gamma$ that is dual to an equal-area triangulation. In this case, all spins are assumed to be equal to $j_0$ and the dimension of the Hilbert space of a node---an equal-area quantum tetrahedron---is simply $\dim\mathcal{H}_n=2j_0+1$. The expectation value of the volume of a node is $v(j_0)=\text{Tr}(\hat{V}_n\, \rho_n)=\frac{1}{2j_0+1}\sum_k v_k$, where $v_k$ are the eigenvalues of the volume operator $\hat{V}_n$ and $\rho_n$ is the maximally mixed state. When expressed in terms of these parameters, the entanglement entropy of a typical state in $\mathcal{H}_{\Gamma j_0}$ is
\begin{equation}
S_A(\textrm{typical})\;\approx\;  \frac{\log(2j_0+1)}{v(j_0)} \, V_A\;-\;\frac{1}{2}\exp\Bigg(\!-\frac{\log(2j_0+1)}{v(j_0)} \;(V_\Gamma-2V_A)\Bigg) \, .
\end{equation}
This is a volume law for the entanglement entropy of a typical state in the Hilbert space $\mathcal{H}_{\Gamma j_0}$. On the other hand, it is known that taking into account the  dynamics---and in particular constraints such as the selection of an eigenstate of an Hamiltonian with local interactions \cite{Vidmar:2017uux,Vidmar:2018rqk,Hackl:2018tyl}---selects states that are non-typical in the Hilbert space and leads to a behavior of the entanglement entropy that deviates qualitatively from Page's law for typical states.

In this paper, we focus on Bell-network states and show that---instead of a volume law---their entanglement entropy obeys an area law,
\begin{equation}
S_A(\textrm{Bell-network})\;\approx\;  \frac{\log(2j_0+1)}{a(j_0)} \, \mathrm{Area}_{A}\;+\dots\,,
\label{eq:area-law}
\end{equation}
where $a(j_0)=8\pi \gamma G \hbar\, \sqrt{j_0(j_0+1)}$ is the area eigenvalue of a boundary link.

\medskip

Bell-network states are defined using squeezed vacuum techniques that enforce prescribed correlations. In particular, Bell-network states \cite{Baytas:2018wjd} have correlations that reduce the general twisted geometry \cite{Freidel:2010aq,Freidel:2010bw} at adjacent nodes to vector geometries \cite{Barrett:2002ur,Dona:2017dvf} by introducing Bell-like correlation in the normals to faces of adjacent polyhedra.
The structure of correlations is well illustrated by the Bell state for two spin-$1/2$ particles,
\begin{equation}
 \ket{\mathcal{B}}=\frac{\ket{\uparrow}_1\ket{\downarrow}_2-\ket{\downarrow}_1 \ket{\uparrow}_2}{\sqrt{2}}\;= \;\sqrt{2}\int \frac{d\vec{n}}{4\pi}\, \ket{\vec{n}}_1\ket{-\vec{n}}_2 \, ,
\end{equation}
which is given by a uniform superposition of back-to-back spins. For a given graph $\Gamma$ and assignment $j_\ell$ of spins, there is a unique Bell-network state in $\mathcal{H}_{\Gamma j_\ell}$, here denoted $|\Gamma,j_\ell,\mathcal{B}\rangle$. Its expansion coefficients in the basis \eqref{eq:s-state} are given by the $SU(2)$ symbol of the graph.

\medskip

Calculating the entanglement entropy $S_A$ of a region $A$ of a Bell-network state is non-trivial. In this paper, we present analytical and numerical methods to compute it. We will work only with finite graphs having a finite number of nodes. The problem is structured in the same way as the standard entanglement entropy computation in many-body quantum systems, where one considers a state (e.g. the ground state of a specific Hamiltonian) and then computes the entanglement entropy for various subsystems \cite{PageTyp,Vidmar:2017uux,Vidmar:2018rqk,Hackl:2018tyl}. In this work, we present a numerical code \cite{code} that, for a given graph, first evaluates the expansion of the Bell-network state on a factorized basis, and then computes the entanglement entropy of various subsystems. We present explicit versions of the code adapted to different graphs and subregions. We consider: the dipole graph, the pentagram graph with subregions containing one or two nodes, and the hexagram graph with subregions containing one or two nodes (either connected or disconnected).

To identify qualitative features of the behavior of the entanglement entropy for any Bell-network state and arbitrary subsystem, we employ analytical methods which provide good approximations under the assumption of uniformly large spins. Under a homogeneous rescaling of the spins of the state $j_\ell\to \lambda j_\ell$, we derive a bound for the leading order in $\lambda\gg 1$ of the entanglement entropy of a region $A$, 
\begin{equation}
\left(\BB - c_A\right) \log \lambda\; \leq \;S_A \;\leq\; \left(\BB-3\right) \log \lambda \, ,
\end{equation}
where $c_A$ is a half-integer and $\BB$ is the number of links that cross the boundary of $A$. In this regime the bound implies an area-law behavior for $S_A$.   We determine also the behavior of the R\'enyi entropy of order $p$ of any Bell-network state and arbitrary region. When compared to our numerical data, we find good agreement within our approximation.

The expectation that entanglement in the degrees of freedom of the gravitational field is a necessary condition for the emergence of a classical spacetime is shared by various approaches to nonperturbative quantum gravity \cite{VanRaamsdonk:2009ar,VanRaamsdonk:2010pw,Bianchi:2012ev,Jacobson:1995ab,Jacobson:2015hqa,Bianchi:2016tmw,Bianchi:2016hmk,Chirco:2017xjb,Cao:2016mst}. The result that Bell-network states satisfy an area law supports the conjecture that entanglement can be used as a probe of semiclassicality in quantum gravity \cite{Bianchi:2012ev}.

\medskip

The paper is structured as follows. In Section \ref{sec:entanglement} we give an elementary introduction to entropic inequalities for the entanglement entropy and the  R\'enyi entropy, together with their application to LQG. In Section \ref{sec:Bell-network} we review the definition of Bell-network states and their relation to vector geometries. In Section \ref{sec:asymptotics} we present the large-spin asymptotic analysis of the R\'enyi entropy and the entanglement entropy for a Bell-network state on a generic graph. In Section \ref{sec:numerics} we present our code and compare the numerical results to our analytical asymptotic formulae for some specific graphs. We conclude with a discussion of our results.

%----------------------------------------------------------------------------
\section{Entanglement entropy and R\'enyi entropy in LQG}
\label{sec:entanglement}
A quantum system composed of two subsystems $A$ and $\bar{A}$ has a Hilbert space given by the tensor product:
\begin{equation}
\label{Hilbertspaces}
\mathcal{H} = \mathcal{H}_{A} \otimes \mathcal{H}_{\bar{A}}\,.
\end{equation}
Given a state $|\psi\rangle$ in $\mathcal{H}$, the reduced density matrix of the subsystem $A$ is defined by the partial trace over its complement $\bar{A}$
\begin{equation}
\label{densitym}
\rho_A = \Tr_{\bar{A}} \;|\psi\rangle\langle\psi| \, .
\end{equation}
The entanglement entropy of the subsystem $A$ is defined as the von Neumann entropy of the reduced density matrix  
\begin{equation}
\label{EEentropy}
S_A=-\Tr\,\big(\rho_A \log\rho_A\big)\, .
\end{equation}
It is also useful to define the R\'enyi entropy of order $p$, defined as
\begin{equation}
\label{Rpentropy}
R_A^{(p)}=-\frac{1}{p-1}\log\Tr\rho_A^p\, ,
\end{equation}
with $p\geq 0$. The limit $p\to 1$ reproduces the entanglement entropy, as can be easily shown:
\begin{align}
\label{limit}
\lim_{p\to1}R_A^{(p)}\;=&
-\lim_{\epsilon\to0}\frac{1}{\epsilon}\log\Tr\big(\rho_A\left(\mathds{1}+\epsilon \log \rho_A\right)\big) \nonumber\\[1em]
=&-\lim_{\epsilon\to0}\frac{1}{\epsilon}\log\left(1+\epsilon\, \Tr \rho_A \log \rho_A\right)=
- \Tr \big(\rho_A \log\rho_A\big)=S_A \, .
\end{align}
On the other hand, the limit $p\to 0$ of the R\'enyi entropy reproduces the maximum entropy  $S^{\max}_A = \log (\mathrm{dim} \mathcal{H}_A)$,
\begin{equation}
\lim_{p\to0}R_A^{(p)}=\log\Tr\lim_{p\to0}\rho_A^p =\log(\mathrm{dim} \mathcal{H}_A)=S^{\max}_A \, .
\end{equation}
The R\'enyi entropies satisfy the useful inequality
\begin{equation}
R_A^{(p)} \leq R_A^{(p')} \quad \text{for}\quad p>p' 
\end{equation}
with the equality corresponding to the maximally mixed state $\rho_A= \mathds{1} / \mathrm{dim} \mathcal{H}_A$ for which $R_A^{(p)} = \log (\mathrm{dim} \mathcal{H}_A)$. Considering the limits $p\to 0,1$, these inequalities provide an upper and a lower bound on the entanglement entropy,
\begin{equation}
\label{inequalities}
R_A^{(p)} \;\leq \;S_A \;\leq\; \log(\mathrm{dim} \mathcal{H}_A)\quad \text{with}\quad p>1 \, .
\end{equation}
This relation is instrumental in our analysis.

\medskip

The structures discussed above apply immediately to states in the LQG Hilbert space $\mathcal{H}_{\Gamma j_\ell}=\mathcal{H}_1\otimes\cdots\otimes \mathcal{H}_N$ with fixed graph $\Gamma$, $N$ nodes, and fixed spins $j_\ell$. In this case, a state of this many-body system is a quantum geometry consisting of $N$ entangled polyhedra. 

\medskip

A similar decomposition of the Hilbert space has been used for the investigation of entanglement in the intertwiner degrees of freedom in \cite{Livine:2017fgq} and in \cite{Feller:2017jqx}, where a class of area-law states with spin $1/2$ is studied. The decomposition discussed in these works differs from the edge-mode decomposition of \cite{Donnelly:2008vx,Donnelly:2011hn,Bodendorfer:2014fua,Hamma:2015xla,Anza:2016fix,Delcamp:2016eya,Gruber:2018lef}, where an enlargement of the Hilbert space is considered. In that case there is a local boundary contribution due to edge modes and a non-local contribution due to intertwiner entanglement (quantum polyhedra). As an example of the distinction between the two definitions of entanglement entropy (intertwiner entanglement vs edge-mode entanglement),  we can consider a spin-network basis state $|\Gamma,j_\ell,i_n\rangle$ and a region $A$. The edge-mode entanglement entropy scales as $\sum
_\ell \log(2j_\ell+1)$, while the intertwiner entanglement entropy is simply zero. We refer to \cite{Casini:2013rba} for a detailed discussion of the relation between different definitions of the entanglement entropy in lattice gauge theory and the related choice of subalgebra of observables.\\

While in this paper we focus on the Hilbert space at fixed spins, $\mathcal{H}_{\Gamma j_\ell}$, the notion of entanglement entropy that we use generalizes immediately to a sum over spins. In fact, the LQG Hilbert space at fixed graph $\Gamma$ does not have a tensor product structure. It is instead given by the direct sum $\mathcal{H}_{\Gamma}=\bigoplus_{j_\ell} \mathcal{H}_{\Gamma j_\ell}$ over spaces at fixed spins. Remarkably, in this case the entanglement entropy can be computed following \cite{Casini:2013rba}. Given a state $\ket{\Gamma, v} \in \mathcal{H}_{\Gamma}$,
\begin{equation}
\ket{\Gamma, v} =\sum_{j_\ell} q_{j_\ell} \ket{\Gamma, j_\ell,v},
\end{equation}
with $\ket{\Gamma, j_\ell,v}\in \mathcal{H}_{\Gamma j_\ell}$, the entanglement entropy of a region $A$ can be computed as
\begin{equation}
\label{eq:EntropyDirectSum}
S_A = - \sum_{j_\ell} p_{j_\ell} \log p_{j_\ell} + \sum_{j_\ell} p_{j_\ell} S_A \left( j_\ell \right)
\end{equation}
where $ p_{j_\ell} = \left|q_{j_\ell}\right|^2 / \sum_{j_\ell} \left|q_{j_\ell}\right|^2$ is the probability of finding the state $\ket{\Gamma, v}$ with definite spins $j_\ell$, and the entropy is the sum of the classical Shannon entropy of the probability distribution $p_{j_\ell}$ and the average entropy at fixed spin.

\section{Bell-network states and vector geometries}
\label{sec:Bell-network}
The LQG Hilbert space at fixed graph $\Gamma$ and spins $j_\ell$ consist of a collection of quantum polyhedra, one for each node. Geometrical quantities like angles, areas, volumes, and shapes of these polyhedra are quantum operators in this Hilbert space. To glue two polyhedra together, we need to impose the matching of the shape of the face shared between the two. At the quantum level, due to the uncertainty relations of shape operators, we cannot require two shape eigenstates to coincide, but we can impose gluing as expectation values. Moreover, we can require correlations between two adjacent quantum polyhedra so that also the fluctuations of the shape of two adjacent faces are correlated. Bell-network states \cite{Baytas:2018wjd,Bianchi:2016tmw,Bianchi:2016hmk} are a specific proposal that uniformly maximizes correlations of all neighboring polyhedra on a given graph. They are given by the formula
\begin{equation}
\label{BNdef}
\ket{\Gamma, j_\ell,\mathcal{B}} =\frac{1}{\sqrt{Z}} \sum_{i_n} \overline{\mathcal{A}_\Gamma \left(j_\ell, i_n \right)} \bigotimes_n \ket{i_n} \, ,
\end{equation}
where $Z$ is a normalization and the amplitude $\mathcal{A}$ is the $SU(2)$-symbol of the graph $\Gamma$, i.e.,
\begin{equation}
\label{BNcoeff}
\mathcal{A}_\Gamma \left(j_\ell, i_n \right) =\sum_{\left\lbrace m\right\rbrace} \prod_n \left[i_n \right]^{m_1 \cdots m_{F_n}} \, .
\end{equation}
Here the intertwining tensors $\left[i_n \right]^{m_1\cdots m_{F_n}}$ are contracted according to the connectivity of the graph $\Gamma$ and $F_n$ is the number of faces of the quantum polyhedron in the node $n$.

These states have an appealing geometrical interpretation: for large spins, they describe a uniform superposition of vector geometries, a collection of polyhedra glued together by requiring that the normals of adjacent faces are back-to-back, even though in general the faces don't have the same shape. This class of geometries plays an essential role in the study of the asymptotic behavior of topological $BF$ $SU(2)$ spin foam vertex amplitudes \cite{Barrett:2009as,Barrett:2009gg,Dona:2017dvf,Han:2011re,Engle:2015zqa,Bianchi:2017hjl}.
Bell-network states have built-in short-range correlations and are expected to satisfy an area law for the entanglement entropy. Proving that the area law arises at the gauge-invariant level is not immediate as it requires us to control correlations in the intertwiner degrees of freedom. Here we present the explicit computations necessary to determine this behavior.\\

While in this paper we focus on Bell-network states at fixed spins, their full definition includes also specific weights for the sum over spins. We review briefly the related construction.

Bell-network states are defined using the formalism of squeezed spin networks, developed in \cite{Bianchi:2016tmw,Bianchi:2016hmk}. The objective is to build entangled states for neighboring quantum polyhedra. 

Given a graph $\Gamma$, the Hilbert space of a link $\ell\in\Gamma$ can be thought of as the Hilbert space of four harmonic oscillators, two at the source and two at the target of the link \cite{Bianchi:2016hmk}. Denoting the creation operators $a^\dagger_s{}^\alpha$ and $a^\dagger_t{}^\alpha$, where $\alpha=1,2$ is a spinor index, we build a Bell state of the link $\ell$ as 
\begin{equation}
\ket{\mathcal{B}, \lambda}_{\ell} = \left( 1- \left| \lambda\right|^2 \right) \exp \left( \lambda \epsilon_{\alpha \beta}a^\dagger_s{}^\alpha a^\dagger_t{}^\beta \right) \ket{0}_s \ket{0}_t \ ,
\end{equation}
where the squeezing paramenter $\lambda$ is a complex number that encodes the average area $A_\ell$ and the average extrinsic angle $\theta_\ell$ associated to the link. The Bell-network state of a full graph $\Gamma$ is then defined as the gauge-invariant projection of the tensor product of link Bell states
\begin{equation}
\ket{\Gamma, \lambda_\ell,\mathcal{B}} = P_\Gamma \otimes_{\ell \in \Gamma} \ket{\mathcal{B}, \lambda_{\ell}}_{\ell} \ .
\end{equation}
The gauge-invariant projection can be implemented using the resolution of the identity in the spin-network basis $P_\Gamma=\sum_{j_\ell, i_n} \ket{\Gamma, j_\ell,i_n} \bra{\Gamma, j_\ell,i_n}$. The result of this projection takes a simple form. We obtain an expression for the graph Bell-network states in terms of a sum over spins
\begin{equation}
\label{BNdefG}
\ket{\Gamma, \lambda_\ell,\mathcal{B}} = \sum_{j_\ell}\left( \prod_\ell \left( 1- \left| \lambda_\ell\right|^2 \right) \lambda_\ell^{2 j_\ell}  \sqrt{2 j_\ell +1} \right) \ket{\Gamma, j_\ell,\mathcal{B}}\,,
\end{equation} 
where $\ket{\Gamma, j_\ell,\mathcal{B}}$ are the states \eqref{BNdef} we focus on in this paper.

Computing the entanglement entropy of $\ket{\Gamma, j_\ell,\mathcal{B}}$ is non-trivial. Once the result of the entanglement entropy at fixed spins is obtained, the entanglement entropy of the full state can be computed using \eqref{eq:EntropyDirectSum} with $q_{j_\ell}=\prod_\ell \left( 1- \left| \lambda_\ell\right|^2 \right) \lambda_\ell^{2 j_\ell}  \sqrt{2 j_\ell +1}$.

\section{Large-spin asymptotic analysis of the entanglement entropy}
\label{sec:asymptotics}

\begin{figure}[t]
\centering
\includegraphics[scale=0.6]{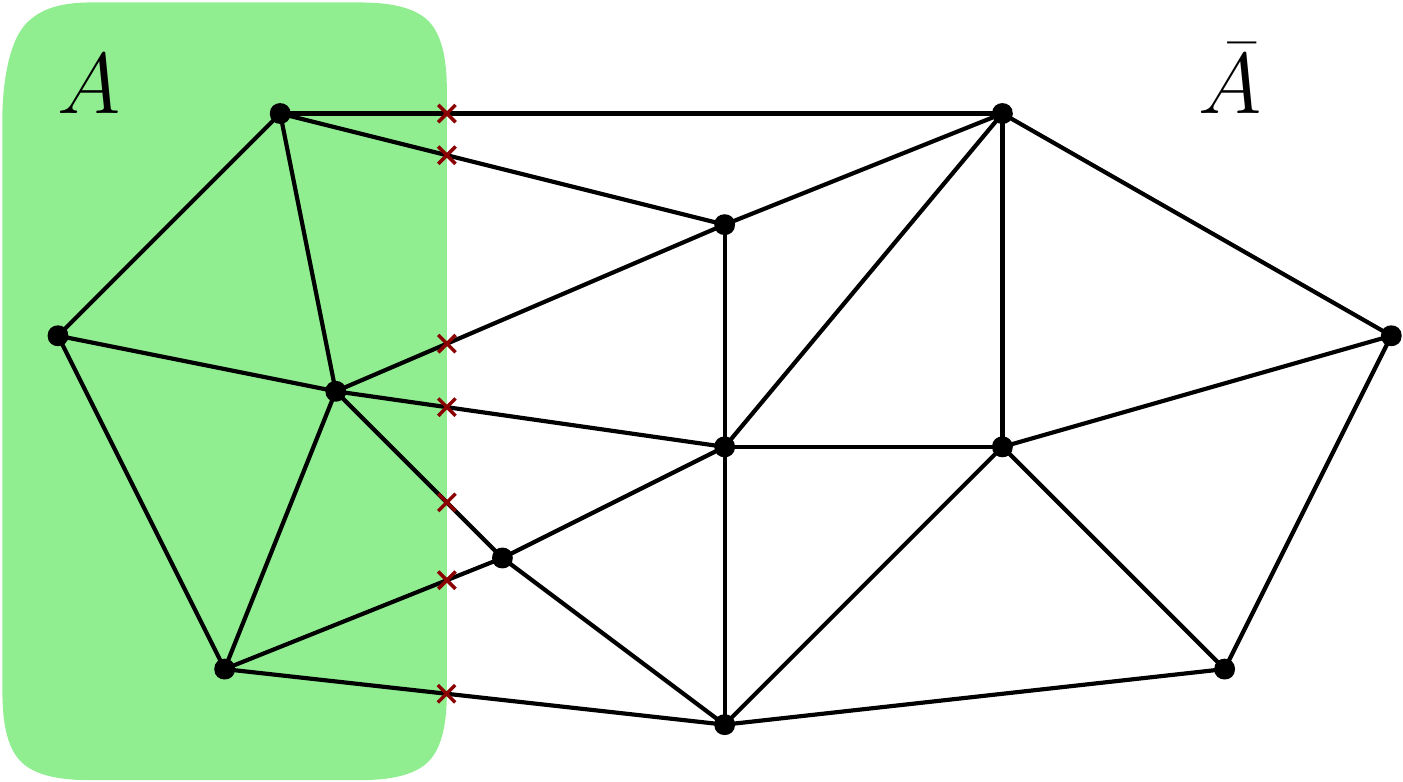}
\caption{\label{fig:exampleSN} Example of a general spin network on a graph $\Gamma$. We shaded in green a region $A$ which contains $N_A=4$ nodes and $\BB=7$ boundary links (marked with a cross). The region $A$ determines a subsystem $\mathcal{H}_{A}=\bigotimes_{n\in A} \mathcal{H}_n$.}
\end{figure}
Given a Bell-network state on a general graph $\Gamma$ with fixed spins $j_\ell$ \eqref{BNdef}, we consider a region $A$ containing a certain number of nodes $N_A$. We denote $\BB$ the number of links crossing the boundary of $A$, see Figure \ref{fig:exampleSN}.  The density matrix of $A$ is defined by tracing away the intertwiners in $\bar{A}$, i.e., 
\begin{align}
\label{densityM}
\rho_A=& \;\Tr_{\bar{A}} \ket{\Gamma,j_\ell,\mathcal{B}}\bra{\Gamma,j_\ell,\mathcal{B}} \;=\; \frac1Z \sum_{i_n, i_n'} M\left(i_n, i'_n\right) \bigotimes_{n\in A} \ket{i_n}\bra{i_n'} \, .
\end{align}
The normalization $Z$  guaranties  that $\Tr \rho =1$. The matrix $M(i_n, i'_n)$ is defined as 
\begin{equation}
M(i_n, i'_n) = \sum_{k_n\in \bar{A}}\overline{\mathcal{A}_\Gamma \left(j_\ell, i_n, k_n \right)}\; \mathcal{A}_\Gamma \left(j_\ell, i_n' ,k_n \right) \, ,
\end{equation}
where the sum is  over the intertwiners $k_n$ associated to the nodes contained in $\bar{A}$, the complement of $A$. The normalization factor is easily expressed in terms of the matrix $M$ as $Z= \Tr M$. The trace of the density matrix raised to a power $p$ can also be expressed in terms of $M$ as $\Tr \rho^p_A= \Tr M^p / \left(\Tr M\right)^p$. This formula allows us to compute the R\'enyi entropies in terms of the matrix $M$:
\begin{equation}
\label{RpentropyM}
R_A^{(p)}=-\frac{1}{p-1}\log\Tr\rho^p_A= -\frac{1}{p-1}\left(\log \Tr M^p -p \log \Tr M \right)\, .
\end{equation}
The ingredients needed in the formula are the traces $\Tr M^p$. For instance, $\Tr M$ can be written in terms of $SU(2)$-symbols as
\begin{equation}
\Tr M=\sum_{i_n,k_n}\overline{\mathcal{A}_\Gamma \left(j_\ell, i_n, k_n \right)} \mathcal{A}_\Gamma \left(j_\ell, i_n ,k_n \right) \,=\,Z \, .
\label{traces}
\end{equation}
Clearly, these quantities can be computed using $SU(2)$ recoupling theory. However, no closed expression is available and one has to resort to numerical or symbolic codes, as the ones discussed in the next section. Here we are interested in the behavior of the R\'enyi entropies under uniform rescaling of the spins. To this end we introduce a reformulation in terms of auxiliary variables familiar in spin foam calculations. This reformulation allows us to estimate the value of the R\'enyi entropies analytically in the large-spin regime using a saddle point approximation.

\medskip

To write the R\'enyi entropies in spinfoam-like variables, we express each of the $p\cdot N$ sums over  intertwiners appearing in $\Tr M^p $ as an integral over $SU(2)$  Wigner matrices,
\begin{equation}
\label{entropy}
\sum_{i}\bar{i}_{m_1\cdots m_F}\, i^{n_1\cdots n_F}=\int_{SU(2)} \mathrm{d}g\, D^{(j_1)\,n_1}_{m_1}\left(g\right)\cdots D^{(j_F)\,n_F}_{m_F}\left(g\right) \, .
\end{equation}
The indices are contracted according to the connectivity of the graph \eqref{BNcoeff} and result into $SU(2)$ characters. In particular, there is one $SU(2)$ character for each link crossing the boundary of $A$, and $p$ characters for each link completely inside or outside  the region $A$. In total, there are $pL-(p-1)\BB$ characters.
By introducing a resolution of the identity in terms of $SU(2)$ coherent states $\ket{j\,\vec{n}}$ \cite{Perelomov:1986tf}, each $SU(2)$ character $\chi^{(j)}= \Tr D^{(j)}$ can be expressed as an integral over a unit vector $\vec{n}\in S^2$,
\begin{equation}
\label{character}
\chi^{(j)}\left(g\right)=\left(2j+1\right)\int_{S^{2}}\mathrm{d}\vec{n}\,\bra{j\,\vec{n}}g\ket{j\,\vec{n}}=\left(2j+1\right)\int_{S^{2}}\mathrm{d}\vec{n}\;\exp \left( 2j\,\log \bra{\tfrac{1}{2}\,\vec{n}} g \ket{\tfrac{1}{2}\,\vec{n}} \right) \, .
\end{equation}
The trace $\Tr M^p$ can then be written as an integral over $SU(2)$ group elements $g_e$ and over unit vectors $\vec{n}_f$ as
\begin{equation}
\label{defF}
\Tr M^p= \int d\vec{n}_f dg_e \,e^{f_p\left(j_\ell,\vec{n}_f,g_e\right)} \, .
\end{equation}
The function $f_p\left(j,\vec{n}_f,g_e\right)$ is linear in $j_\ell$ and can be determined using diagrammatic techniques as illustrated for a specific example in the Appendix \ref{app:Example}.

In the large spin limit, the integral \eqref{defF} can be evaluated using a saddle point approximation. Under a uniform rescaling of all the spins $j_\ell \to \lambda j_\ell$, the  function scales as $f_p\left(j_\ell,\vec{n}_f,g_e\right)\to \lambda f_p\left(j_\ell,\vec{n}_f,g_e\right)$, and the leading order in $\lambda$ of the logarithm of the trace $\Tr M^p$ is given by:
\begin{equation}
\log\left( \Tr M^p\right)=\left( \#S^2\textit{integrals}-\frac{1}{2}\#\textit{Hessian}\left(f_p\right) \right)\log\lambda + O(1) \, .
\end{equation}
The first term, the number of integrals over coherent states, is due to the dimensional factor $2j+1$ in \eqref{character}. The second term, $\#\textit{Hessian}$, is the rank of the Hessian of $f_p$ and  it can be expressed in terms of the number of $SU(2)$ integrals, the number of $S^2$ integrals, the number of symmetries of $f_p$ and the dimension of the space of solutions of the system of saddle point equations (denoted $\#\textit{space}\,\textit{of}\,\textit{solutions}$). If multiple saddle points exist, the dominant saddle point is characterized by the largest space of solutions. Therefore, 
\begin{align}
\#Hessian\left(f_p\right)\;=\;&+ 3\times\#SU(2)\textit{integrals} \\
&+2\times\#S^2\,\textit{integrals}\\
&-\#\textit{symmetries}\\
&-\#\textit{space}\,\textit{of}\,\textit{solutions}\left(f_p\right) \, .
\end{align}
The symmetries of $f_p$ are related to right and left $SU(2)$ multiplication of the integration group elements $g_e$. It can be shown that the function $f_p$ always has $2p$ SU(2) symmetries, resulting in $\#\textit{symmetries}=3 \times 2p$. As already discussed, the number of $SU(2)$ integrals are $p$ times the total number $N$ of nodes of the graph $\Gamma$. Therefore,
\begin{equation}
\label{finaltrMp}
\log\left( \Tr M^p\right)=-\frac{1}{2}\big(3pN - 6p -\#\textit{space}\,\textit{of}\,\textit{solutions}\left(f_p\right)\big) \log \lambda + O(1) \, .
\end{equation}
The number $\#\textit{space}\,\textit{of}\,\textit{solutions}$ is given by the total number of independent variables (2 per unit vector), minus the number of independent critical-point equations and a global rotation.
In the case $p=1$, the number of independent critical-point equations coming from $f_1$ is $3(N-1)$ (i.e., $N-1$ vectorial equations), resulting in $\#\textit{space}\,\textit{of}\,\textit{solutions} \left( f_1\right)=2L-3 - 3(N-1)=2L-3N$. This number appears in  \cite{Dona:2017dvf} where it is derived in the context of the asymptotic analysis of spin foam amplitudes of topological theories.

In the general case, denoting
\begin{equation}
C^{(p)}_A=\#\textit{redundant critical-point equations of}\; f_p\, ,
\end{equation}
we find that the number of independent critical-point equations of $f_p$ is $3pN-C^{(p)}_A-3$, corresponding to $pN$ total vectorial equations minus the number of redundant equations ($C^{(p)}_A$) and a global rotation.

The number $C^{(p)}_A$ can be computed case-by-case for a given graph $\Gamma$, region $A$ and power $p$. However, a general closed formula is not available. We note that $C^{(p)}_A$ has two important properties: $C^{(p)}_A$ is an integer and is bounded from below by 
\begin{equation}
C^{(p)}_A\geq 6\left(p-1\right)\, ,
\end{equation}
which is the number of equations that are redundant because of symmetries\footnote{We note that there are $2p$ symmetries $f_p\left(j_\ell,\vec{n}_f,\tilde g_e\right)=f_p\left(j_\ell,\vec{n}_f, g_e\right)$ with $\tilde{g}_e=h g_e$ or $\tilde{g}_e=g_e h$ for some $e$, with $h\in SU(2)$. For $h$ close to the identity, these symmetries result in a linear constraint $\delta_h f_p\left(j_\ell,\vec{n}_f,\tilde g_e\right)=0$ on the critical-point equations.  }.
The resulting expression is $\#\textit{space}\,\textit{of}\,\textit{solutions} \left( f_p \right)\;=\;2(pL-(p-1)\BB) - 3 - 3pN+C^{(p)}_A+3$.\\

In summary, at the leading order in $\lambda$, the R\'enyi entropy of a Bell-network state is
\begin{align}
\nonumber
R_A^{(p)}\;=\;&-\frac{1}{2\left(p-1\right)}\left(2(pL-(p-1)\BB) - 3pN+C^{(p)}_A - p \left( 2L-3N \right)\right) \log \lambda + O(1)=\\[1em]
=\;&\Bigg(\BB -\frac{C^{(p)}_A}{2\left(p-1\right)}\Bigg) \log \lambda + O(1) \, .
\label{eq:renyientropyfinal}
\end{align}
Using the properties of R\'enyi entropy, we can characterize the dependence of $C^{(p)}_A$ on the order $p$. From the set of inequalities \eqref{inequalities} we find 
\begin{equation}
\frac{C^{(p)}_A}{2\left(p-1\right)}\leq \frac{C^{(p+1)}_A}{2p}\quad \Rightarrow \quad 
C^{(p)}_A<\frac{p}{p-1} C^{(p)}_A\leq C^{(p+1)}_A \, .
\end{equation}
The parameter $C^{(p)}_A$ is monotonically increasing in $p$ and, since $R_A^{(p)}$ is positive, for each $p>1$ is bounded from above by $2\left(p-1\right)\left(\BB-3\right)$. \\

As discussed in Section \ref{sec:entanglement} the entanglement entropy can be obtained from the limit of $p\to1$ of the R\'enyi entropy of order $p$. At the leading order in $\lambda\gg 1$,
\begin{equation}
\label{Slimit}
S_A= \lim_{p\to 1} R^{(p)}_A \approx \lim_{p\to 1} \left(\BB - \frac{C^{(p)}_A}{2\left(p-1\right)}\right) \log \lambda =\left(\BB  - c_A\right) \log \lambda 
\end{equation}
where we denoted as $c_A$ the limit $\lim_{p\to 1} \frac{C^{(p)}_A}{2\left(p-1\right)}$. This limit exists and is finite since the entanglement entropy of a system with a finite number of degrees of freedom is a well-defined quantity. From the properties of $C^{(p)}_A$, we also find that 
\begin{equation}
R^{(p)}_A \approx \left(\BB - \frac{C^{(p)}_A}{2\left(p-1\right)}\right) \log \lambda \;\leq\; \left(\BB-3\right) \log \lambda \,.
\end{equation}
 Note that the inequalities here are understood as asymptotic statements holding at the leading order in $\lambda\gg 1$. Combining the limit of this inequality (that is guaranteed to exist by the monotonicity of $C^{(p)}_A$) with the most strict of the inequality in \eqref{inequalities} we can determine that at the leading order in $\lambda$ the entanglement entropy of a region $A$ of a Bell-network state is bounded from below and above by
\begin{equation}
\label{entropy_bounds}
\left(\BB - \frac{C^{(2)}_A}{2}\right) \log \lambda\; \leq\; S_A\; \leq \;\left(\BB-3\right) \log \lambda \, .
\end{equation}

The explicit computation of $C^{(p)}_A$ requires the analysis of critical-points equations for a given graph $\Gamma$ and subsystem $A$. However, in the special case 
of a subsystem $A$ containing one single node, we can prove that  $C^{(p)}_A=6(p-1)$, which is independent of the number of boundary links. In this case, at the leading order, the R\'enyi entropy of order $p$ is independent of $p$, and \eqref{Slimit} implies that $S_A = \left(\BB-3\right) \log \lambda$. This is an area law.

In general, while we don't have a closed formula for $c_A$, in order to show that an area law arises we only need that $c_A$ does not grow with $\BB$. If this is the case, then---for a region with a large number of boundary links---we obtain an area law.

%----------------------------------------------------------------------------

\section{Large-spin numerical analysis of the entanglement entropy}
\label{sec:numerics}
We provide a numerical code to compute the entanglement entropy and the R\'enyi entropy of any order of the density matrix of a Bell-network state restricted to a region $A$. In this section, we provide three explicit examples with Bell-network states defined on different graphs: the dipole graph, the pentagram graph, and the hexagram graph. We consider subsystems defined by regions $A_i$ containing one or two nodes. The code for the pentagram and the hexagram graph are available in \cite{code}.
\begin{figure}[t]
    \centering
    \begin{subfigure}[t]{0.30\textwidth}
        \centering
        \includegraphics[scale=0.4]{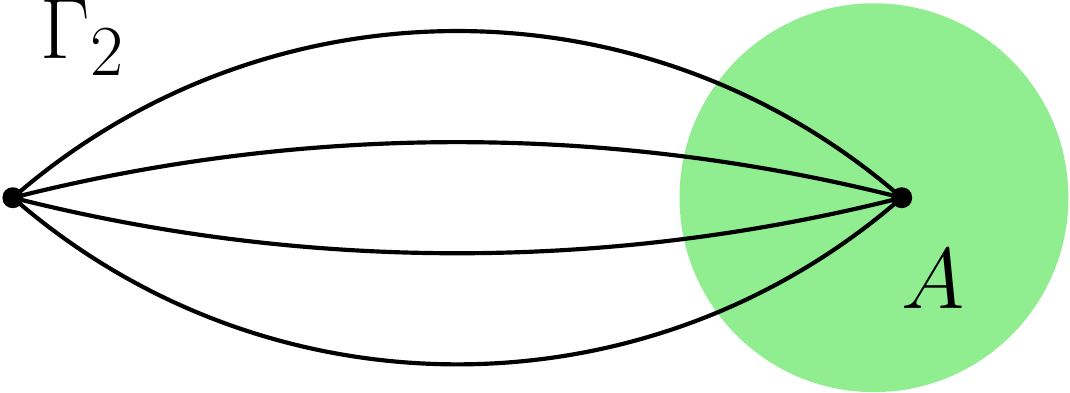}
        \caption{\label{graphdipole}Dipole Graph: one node subsystem $A$}
    \end{subfigure}%
    \hspace{1.5em}
    \begin{subfigure}[t]{0.30\textwidth}
        \centering
		\includegraphics[scale=0.4]{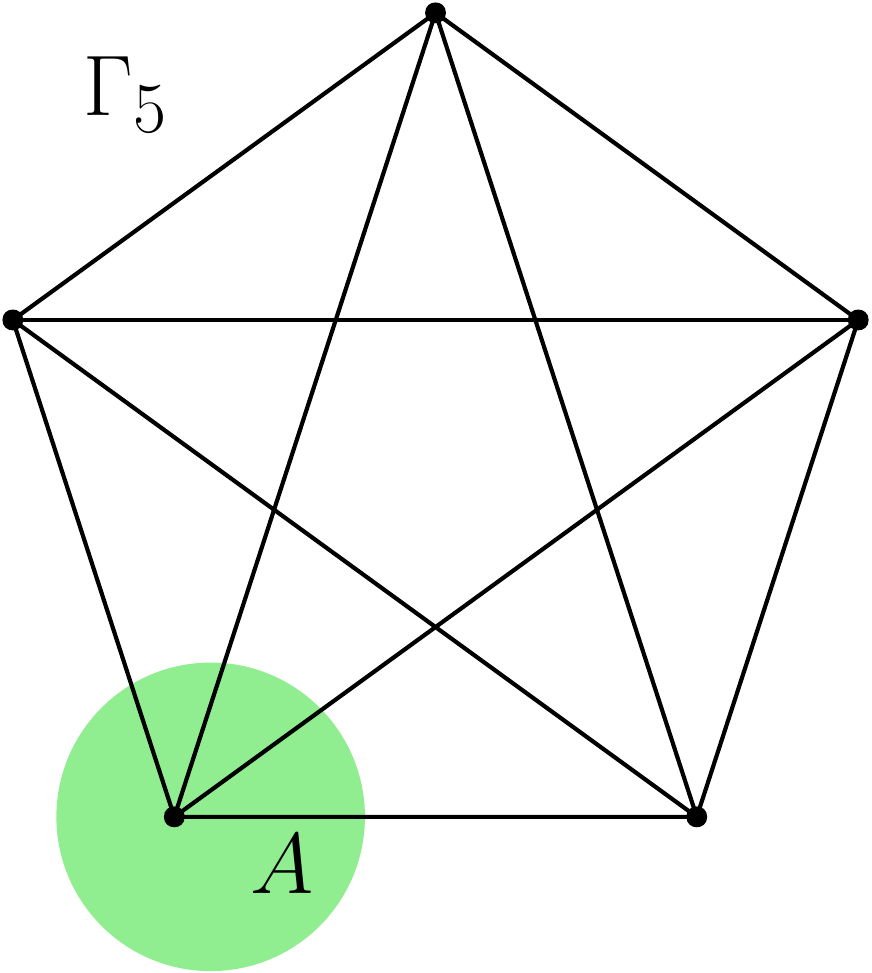}
        \caption{\label{graphpenta1}Pentagram Graph: one node subsystem $A$}
    \end{subfigure}
    \hspace{1.5em}
    \begin{subfigure}[t]{0.30\textwidth}
        \centering
		\includegraphics[scale=0.4]{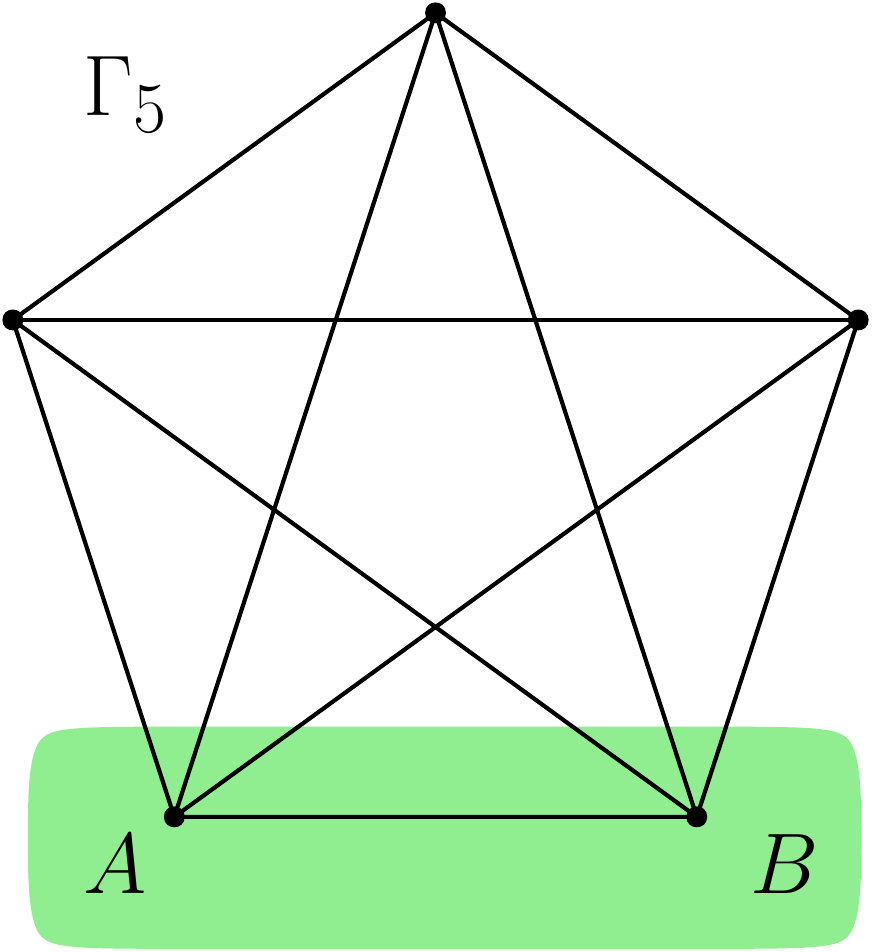}
        \caption{\label{graphpenta2}Pentagram Graph: two node subsystem $AB$}
    \end{subfigure}\\[3em]
        \centering
    \begin{subfigure}[t]{0.30\textwidth}
        \centering
        \includegraphics[scale=0.4]{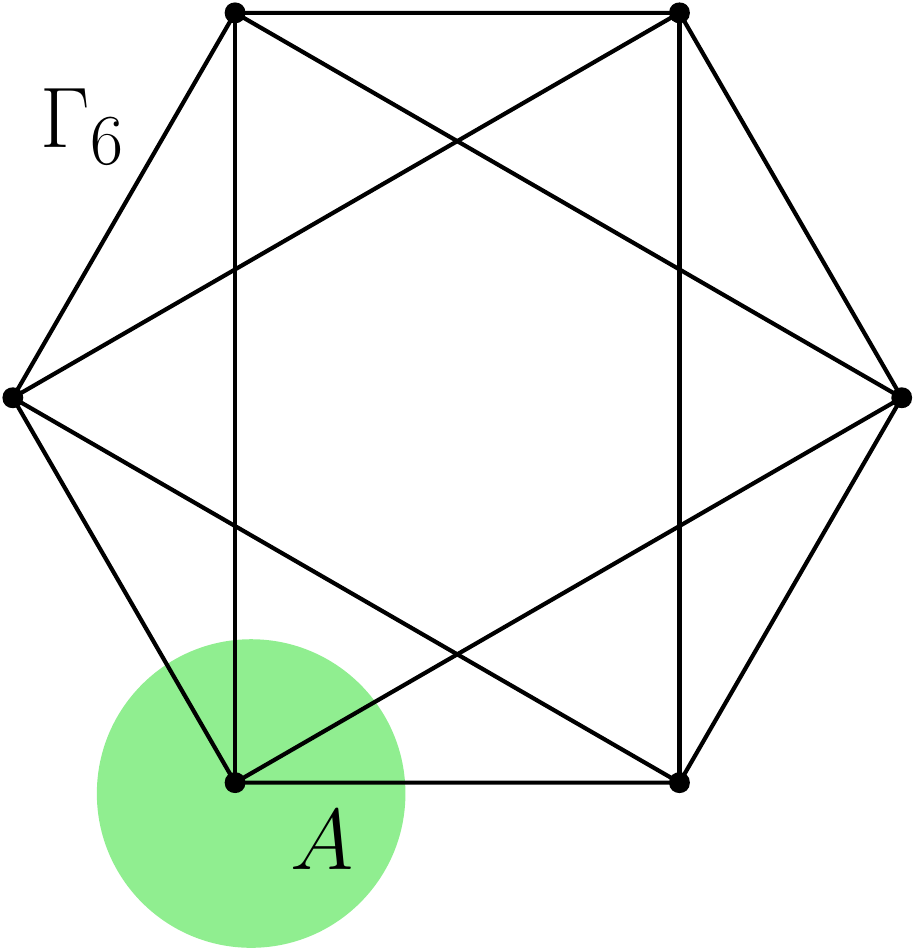}
        \caption{\label{graphhexa1}Hexagram Graph: one node subsystem $A$}
    \end{subfigure}%
    \hspace{1.5em}
    \begin{subfigure}[t]{0.30\textwidth}
        \centering
		\includegraphics[scale=0.4]{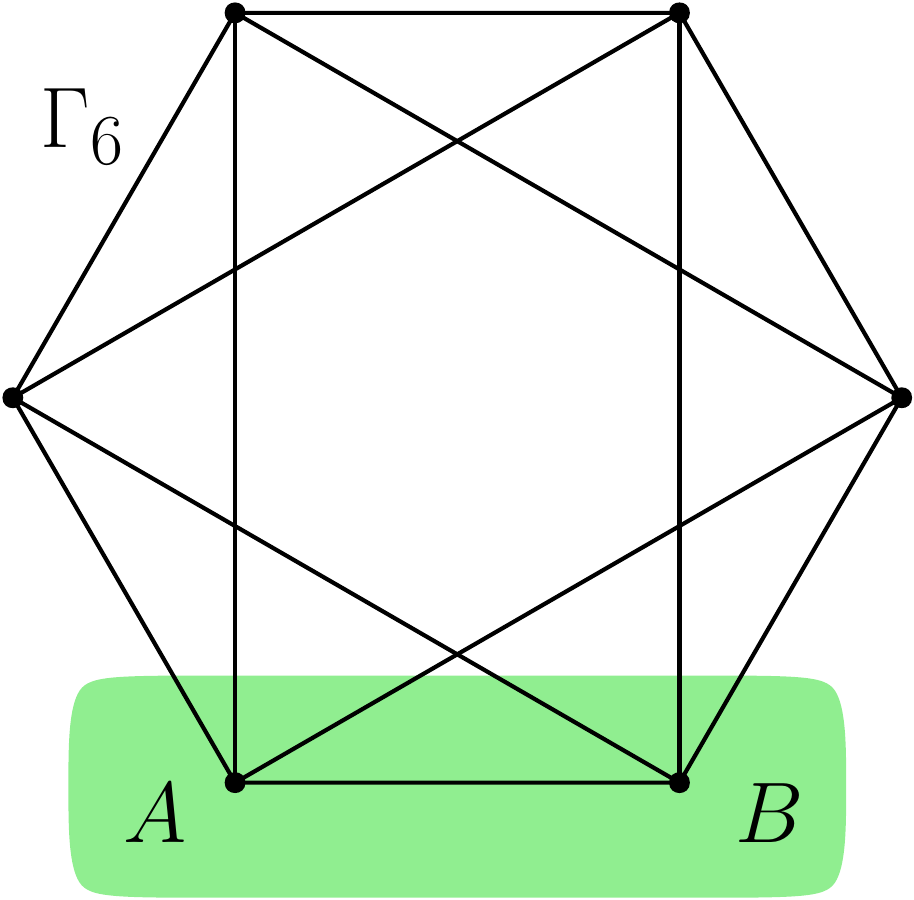}
        \caption{\label{graphhexa2}Hexagram Graph: two connected nodes subsystem $AB$}
    \end{subfigure}
    \hspace{1.5em}
    \begin{subfigure}[t]{0.30\textwidth}
        \centering
		\includegraphics[scale=0.4]{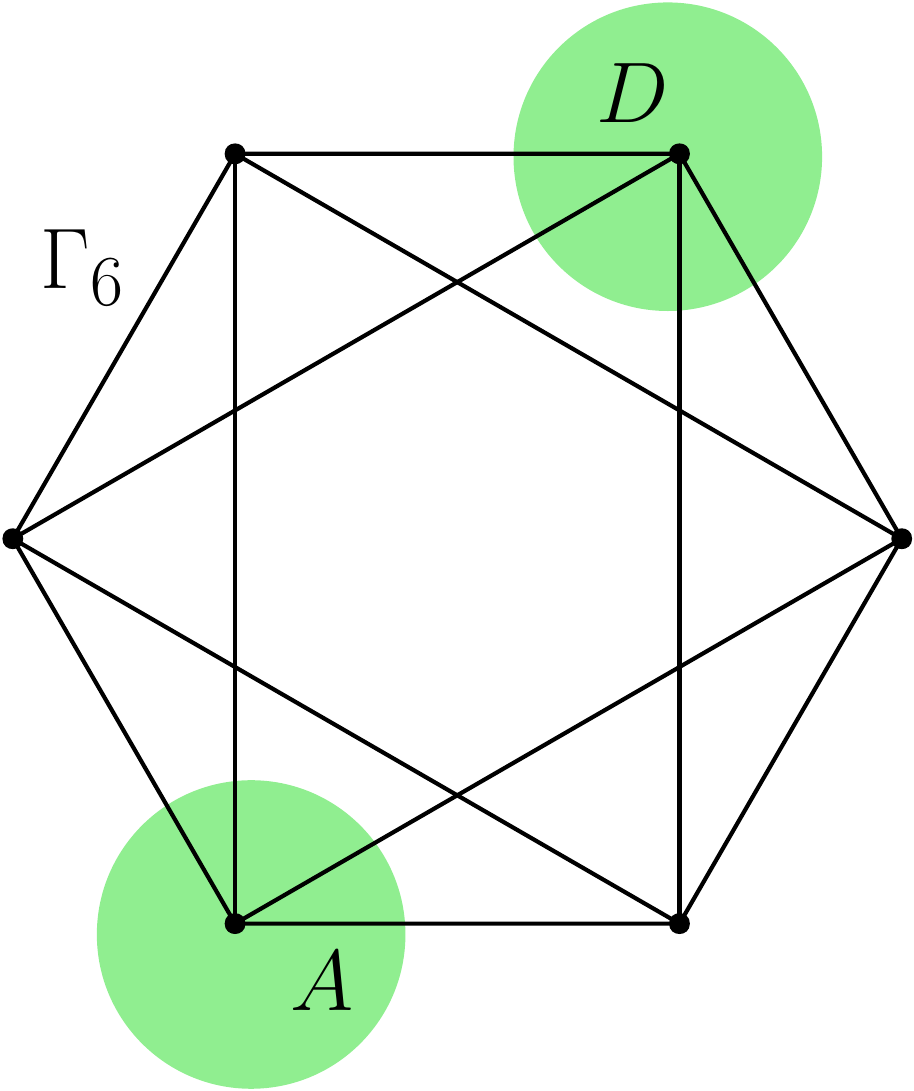}
        \caption{\label{graphhexa3}Hexagram Graph: two disconnected nodes subsystem $AD$}
    \end{subfigure}\\[1em]
\caption{\label{fig:graphs} \small{\emph{Graphs and regions considered in our numerical analysis.}}}
\end{figure}

The algorithm implemented in our  {\tt C} code  is illustrated in the panel below (See Algorithm \ref{numericalcode}). The key ingredient of the calculation is the precomputation of all the $\left\lbrace 6j\right\rbrace$ and $\left\lbrace 9j\right\rbrace$ symbols needed for the evaluation of the symbol of the graph. To efficiently perform this task we employ the \texttt{wigxjpf} library and its extension \texttt{fastwigxj} developed in \cite{Johansson:2015cca} and previously employed in LQG computations \cite{Dona:2017dvf,Dona:2018nev}. 
\begin{algorithm}[H]
\caption{Numerical algorithm for the evaluation of the Bell-network state entropy}\label{numericalcode}
\begin{algorithmic}[1]
\State Precompute the $\{6j\}$ symbols with \texttt{wigxjpf} if needed
\State Precompute the $\{9j\}$ symbols with \texttt{wigxjpf} if needed
\State Load the symbol tables in the memory
\For{each $j_\ell$}
\For{each $i_n$}
\State Assemble the matrix $(\mathcal{A}_\Gamma)_{ik}=\mathcal{A}_\Gamma\left(j_\ell,i_n,k_n\right)$ with the intertwiners $i_n$, $k_n$ in $A$, $\bar{A}$
\EndFor
\State Compute the matrix $M_{ii'}=\left(\mathcal{A}_\Gamma^T \cdot \mathcal{A}_\Gamma\right)_{ii'}$
\State Normalize it to obtain the density matrix $(\rho_A)_{ii'}=M_{ii'}/Tr\left(M\right)$
\State Find its eigenvalues $\rho_A \to \nu_i$
\State Compute the entanglement entropy $S_A=-\sum_i \nu_i \log \nu_i$
\State Compute the R\'enyi entropy $R^{(p)}_A=-\log \sum_i \nu_i^p$
\EndFor
\end{algorithmic}
\end{algorithm}

The range of applicability of our numerical code is limited to spins up to $O(20)$ because of two factors. First, we need a considerable amount of RAM to keep accessible all $\{6j\}$ and $\{9j\}$ symbols required in the computation. For example, the computation of any symbol with spins up to $25$ (both integers and half-integers)  requires approximately $15\,$GB of available RAM, while its extension to spin $30$ requires approximately $40\,$GB of available RAM, not commonly available on ordinary Laptop Computers. This obstacle can be possibly circumvented performing a selection of the symbols prepared and loaded by \texttt{fastwigxj}. At the present stage, we are not selecting symbols and therefore we need to load all of them in the memory.
Second, we use an array of double-precision floating-point numbers to store the symbols. Compilers generally limit the size of this array to the amount of available RAM. For example, the array of symbols for the hexagram graph occupies $(2j+1)^6 \cdot 8$ Byte. We executed our code on a machine with $16\,$GB of RAM, therefore our maximum spin was limited by a  hard cutoff at about $j_{max}\approx\frac{1}{2} \left(\sqrt[6]{(16\,\mathrm{GB})/(8\,\mathrm{Byte})}-1\right)\approx 17$.\\

In the following, we report the numerical computation of the entanglement entropy and the R\'enyi entropy of order two for a set of specific cases.

We set all spins equal $j_\ell = \frac{1}{2}$ and rescale them with a parameter $\lambda$ so that $j_\ell\to \lambda/2$. We will compare the numerical results to the analytical bounds derived in \eqref{entropy_bounds}. We note that in all cases considered, the number of nodes in the region $A$ is small. As a result, the bound $R^{(0)}_A= \log \mathrm{dim} \mathcal{H}_A$ on the entanglement entropy $S_A$ is tighter than the bound $\left(\BB-3\right) \log \lambda$.

\subsection{The dipole graph}
The dipole Bell-network state takes a simple form, computed explicitly in \cite{Baytas:2018wjd}. In this case, analytical computations of the entropy are possible and useful for checking some of the properties derived before. We consider a dipole graph  with four links ($\Gamma_2$, see Figure \ref{graphdipole}):
\begin{equation}
\ket{\Gamma_2, j_\ell,\mathcal{B}} = \frac{1}{\sqrt{\mathrm{dim}\mathcal{H}_1}} \,\sum_{i \in \mathcal{H}_1} \;\ket{i}_1\ket{i}_2 \, ,
\end{equation}
where $\mathcal{H}_1$ is the intertwiner space of a node. The two intertwiners are maximally entangled. Choosing a region  $A$ that contains a single node, the reduced density matrix is:
\begin{equation}
\rho_A = \Tr_2 \left( \ket{\Gamma_2, j_\ell,\mathcal{B}}\bra{\Gamma_2, j_\ell,\mathcal{B}} \right) = \frac{\mathds{1}}{\mathrm{dim}\mathcal{H}_1} \, .
\end{equation}
The resulting state is maximally mixed, and all the entropies are maximal and equal to 
\begin{equation}
\label{exactRenyidipole}
S_A=R^{(p)}_A= \log \mathrm{dim} \mathcal{H}_1 \,.
\end{equation}
We can verify the asymptotic formula \eqref{eq:renyientropyfinal} from the exact computation in the limit of large spins. In the 4-valent case $\BB=4$ the asymptotic estimate reduces to $S_A =R^{(p)}_A= \log \lambda$ which can also be obtained from the uniform rescaling of \eqref{exactRenyidipole}. A similar conclusion can be reached for a dipole graph with an arbitrary number of links greater or equal to three.

%----------------------------------------------------------------------------

\begin{figure}[t]
\centering
\includegraphics[width=0.9\textwidth]{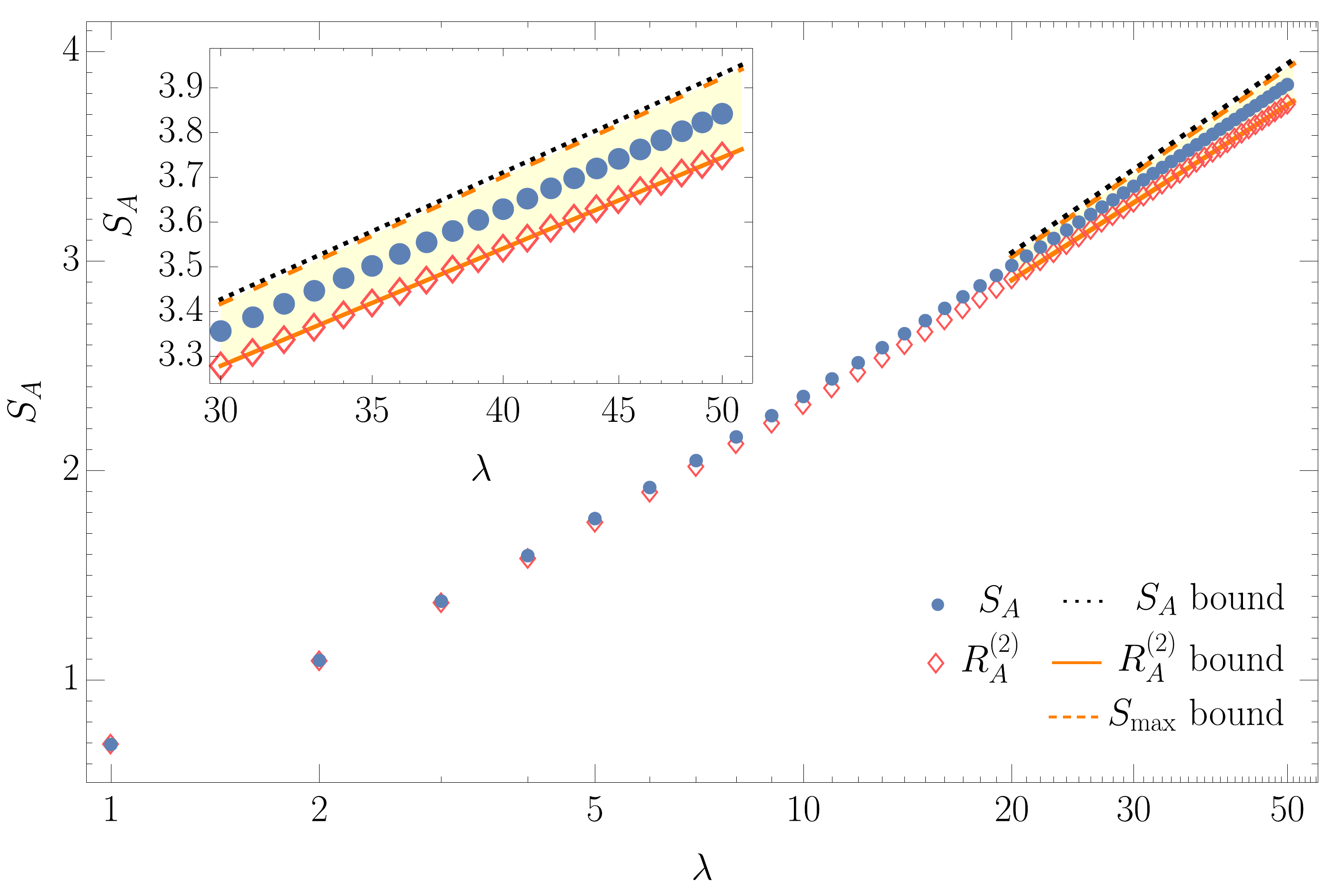}
\caption{[\emph{Graph $\Gamma_5$, entropy of a single node}]. The figure shows our numerical results on the Bell-network state with pentagram graph $\Gamma_5$ and equal spins $j_\ell = \lambda/2$. The entanglement entropy $S_{A}$ is denoted by blue dots, the R\'enyi entropy $R^{(2)}_{A}$ by red diamonds. The lower bound asymptotic estimate is shown as a solid orange line (the $O(1)$ is fitted using the numerical data). The upper bound estimate given by the maximal entropy is shown as a dashed orange line. We show also the bounds  \eqref{entropy_bounds} as a yellow band. The inset shows the $20$ data points with the largest spins.}
\label{fig:penta1n}
\end{figure}

\begin{figure}[t]
\centering
\includegraphics[width=0.9\textwidth]{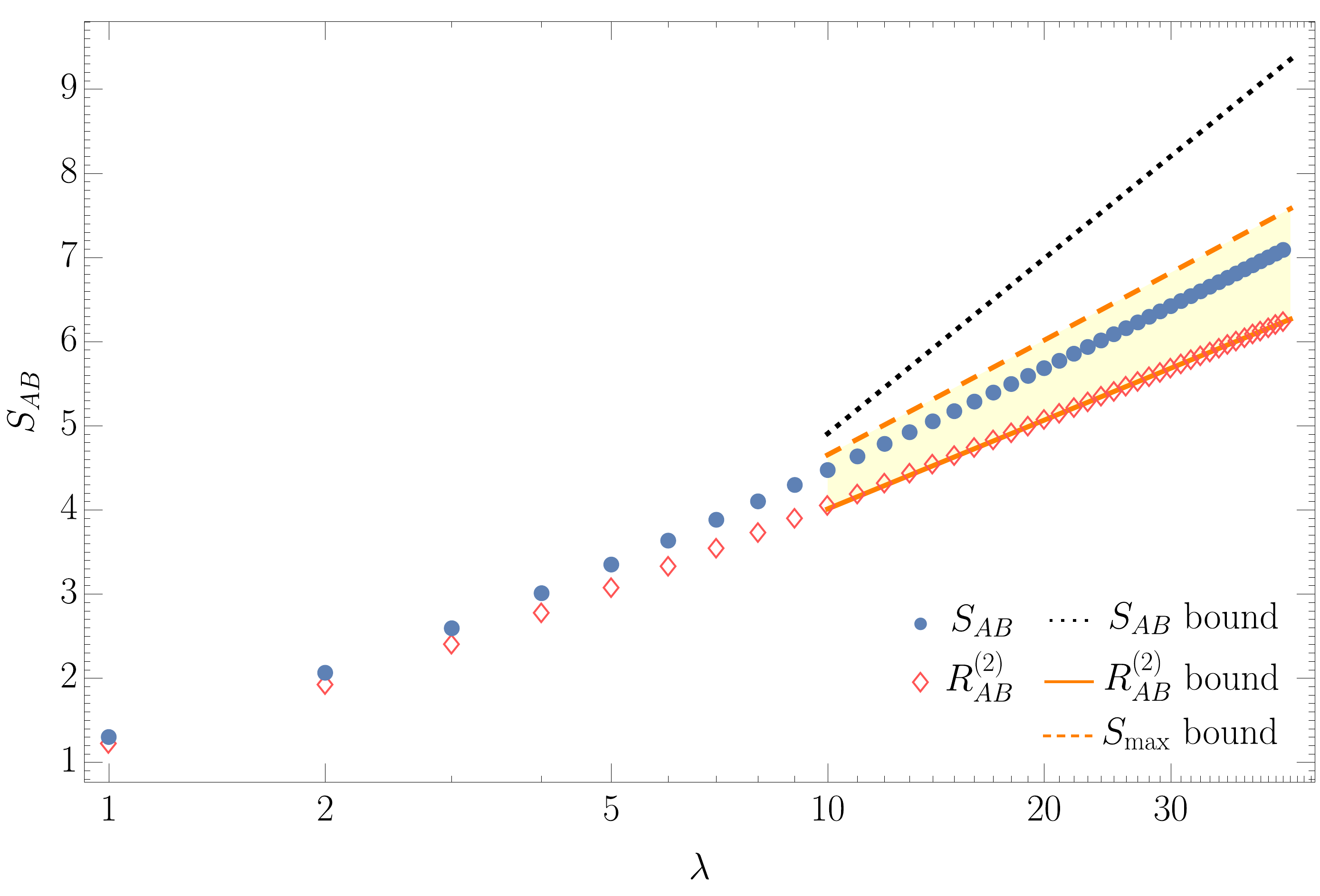}
\caption{[\emph{Graph $\Gamma_5$, entropy of two connected nodes}]. The figure shows our numerical results on the Bell-network state with pentagram graph $\Gamma_5$ and equal spins $j_\ell = \lambda/2$. The entanglement entropy $S_{AB}$ is denoted by blue dots, the R\'enyi entropy $R^{(2)}_{AB}$ by red diamonds. The lower bound asymptotic estimate is shown as a solid orange line (the $O(1)$ is fitted using the numerical data). The upper bound estimate given by the maximal entropy is shown as a dashed orange line. We show also the bounds  \eqref{entropy_bounds} as a yellow band. Using the $10$ largest-spin data points, the entanglement entropy $S_{AB}$ is  fitted by $a \log(\lambda) + b + c \lambda^{-1}$ on the last 10 data points obtaining $a\approx 1.94$, $b\approx -0.30$ and $c\approx 3.19$.}
\label{fig:penta2n}  
\end{figure}

\subsection{The pentagram graph}
The pentagram Bell-network state (see the graph $\Gamma_5$ in Figures \ref{graphpenta1}, and \ref{graphpenta2}) is a superposition of the intertwiner states weighted by the symbol of the graph, the $\left\lbrace 15j\right\rbrace$ symbol,
\begin{equation}
\ket{\Gamma_5, j_\ell,\mathcal{B}} = \frac{1}{\sqrt{Z}}\sum_{i_n} \;\overline{\left\lbrace 15j\right\rbrace \left( j_\ell, i_n\right)}\; \ket{i_1}\ket{i_2}\ket{i_3}\ket{i_4}\ket{i_5} \,.
\end{equation}

\paragraph{One node subsystem of $\Gamma_5$.} Choosing a region $A$ containing a single node, our asymptotic formula reduces to 
\begin{equation}
S_{A} \approx R^{(p)}_{A}\approx \log \lambda
\end{equation}
at the leading order. Note that $S_{A}$ and $R^{(p)}_{A}$ can differ by $O(1)$ terms. We used our code to compute the entanglement entropy and R\'enyi entropy of order two for all equal spins $j_\ell = \lambda/2$ and the scale parameter up to $\lambda\leq 50$. We report the numerical results in Figure \ref{fig:penta1n}. The plot shows clearly that $S_{A}$, $R^{(2)}_{A}$ and the maximal entropy $R^{(0)}_{A} = \log \left( \lambda +1 \right) \approx \log \lambda + O\left(\lambda^{-1} \right)$ differ only by a constant contribution. We interpret this $O(1)$ difference as an indication that the restriction of the Bell-network state is not maximally mixed and the state is not typical in the Hilbert space.

\paragraph{Two nodes subsystem of $\Gamma_5$.} If we choose a subsystem $AB$ containing two nodes, our asymptotic estimates reduce to 
\begin{equation}
R^{(2)}_{AB}= \frac{3}{2}\log \lambda+ O(1)
\end{equation}
and to an asymptotic band for the entanglement entropy given by 
\begin{equation}
\frac{3}{2}\log \lambda \leq S_{AB} \leq 3\log \lambda\,.
\end{equation}
For this specific configuration, the bound from above given by the maximal entropy $S_{AB} \leq 2\log \lambda$ is tighter. The results of the computation of the entanglement entropy and R\'enyi entropy of order two for the case of all equal spins $j_\ell = \frac{\lambda}{2}$ and the scale parameter up to $\lambda\leq 44$ are reported in Figure \ref{fig:penta2n}.

%----------------------------------------------------------------------------

\subsection{The hexagram graph}
The hexagram Bell-network state ($\Gamma_6$, see Figures \ref{graphhexa1}, \ref{graphhexa2}, and \ref{graphhexa3}) is a superposition of the intertwiner states weighted by the symbol of the graph, the $\left\lbrace 18j\right\rbrace$ symbol,
\begin{equation}
\ket{\Gamma_6, j_\ell,\mathcal{B}} = \frac{1}{\sqrt{Z}}\sum_{i_n}\; \overline{\left\lbrace 18j\right\rbrace \left( j_\ell, i_n\right)} \;\ket{i_1}\ket{i_2}\ket{i_3}\ket{i_4}\ket{i_5}\ket{i_6} \,.
\end{equation}

\begin{figure}
\centering
\includegraphics[width=0.9\textwidth]{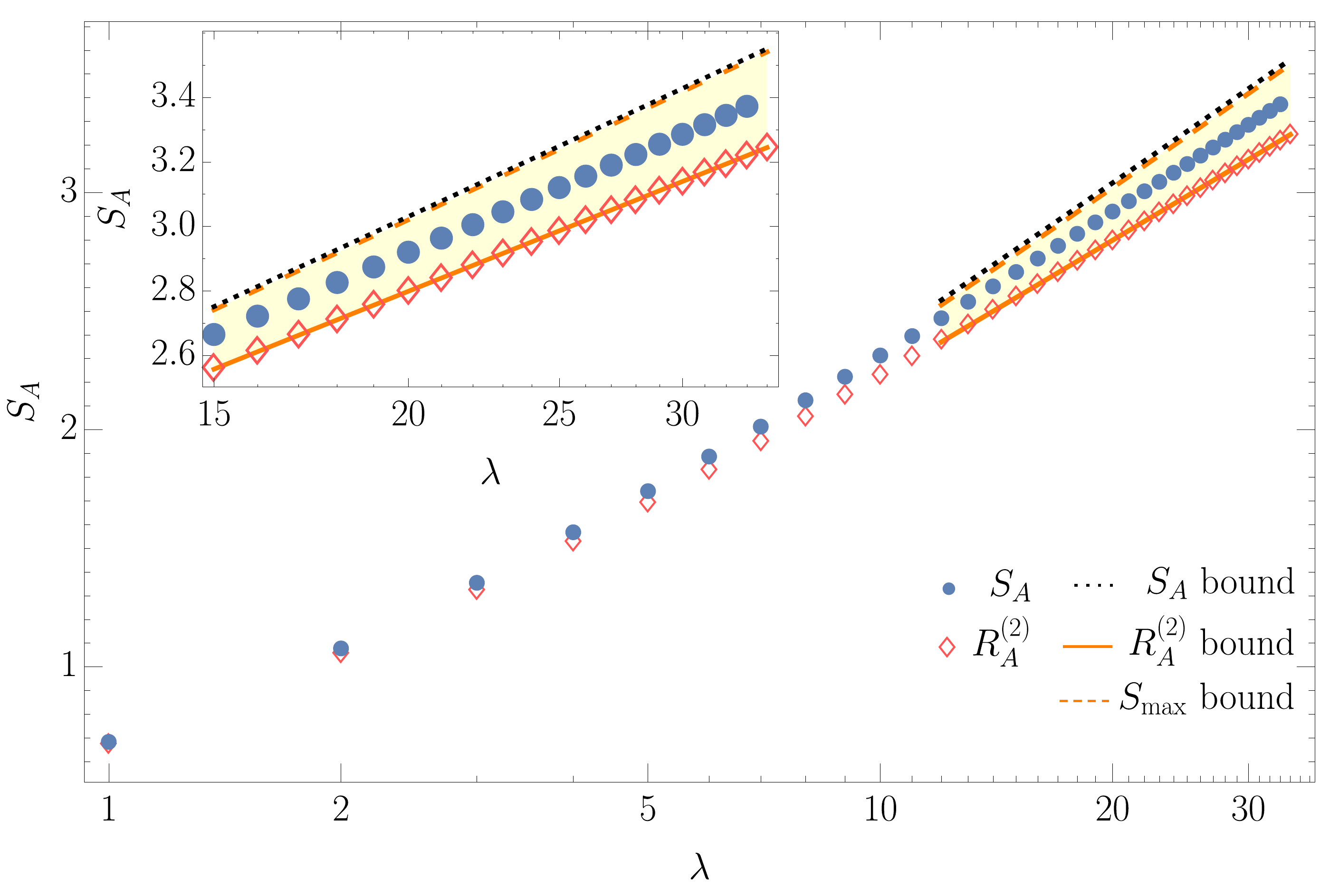}
\caption{[\emph{Graph $\Gamma_6$, entropy of a single node}]. The figure shows our numerical results on the Bell-network state with hexagram graph $\Gamma_6$ and equal spins $j_\ell = \lambda/2$. The entanglement entropy $S_{A}$ is denoted by blue dots, the R\'enyi entropy $R^{(2)}_{A}$ by red diamonds. The lower bound asymptotic estimate is shown as a solid orange line (the $O(1)$ term is fitted using the numerical data). The upper bound estimate given by the maximal entropy is shown as a dashed orange line. We show also the bounds  \eqref{entropy_bounds} as a yellow band. The inset shows the $20$ data points with the largest spins.}
\label{fig:hexa1n}  
\end{figure}

\begin{figure}
\centering
\includegraphics[width=0.9\textwidth]{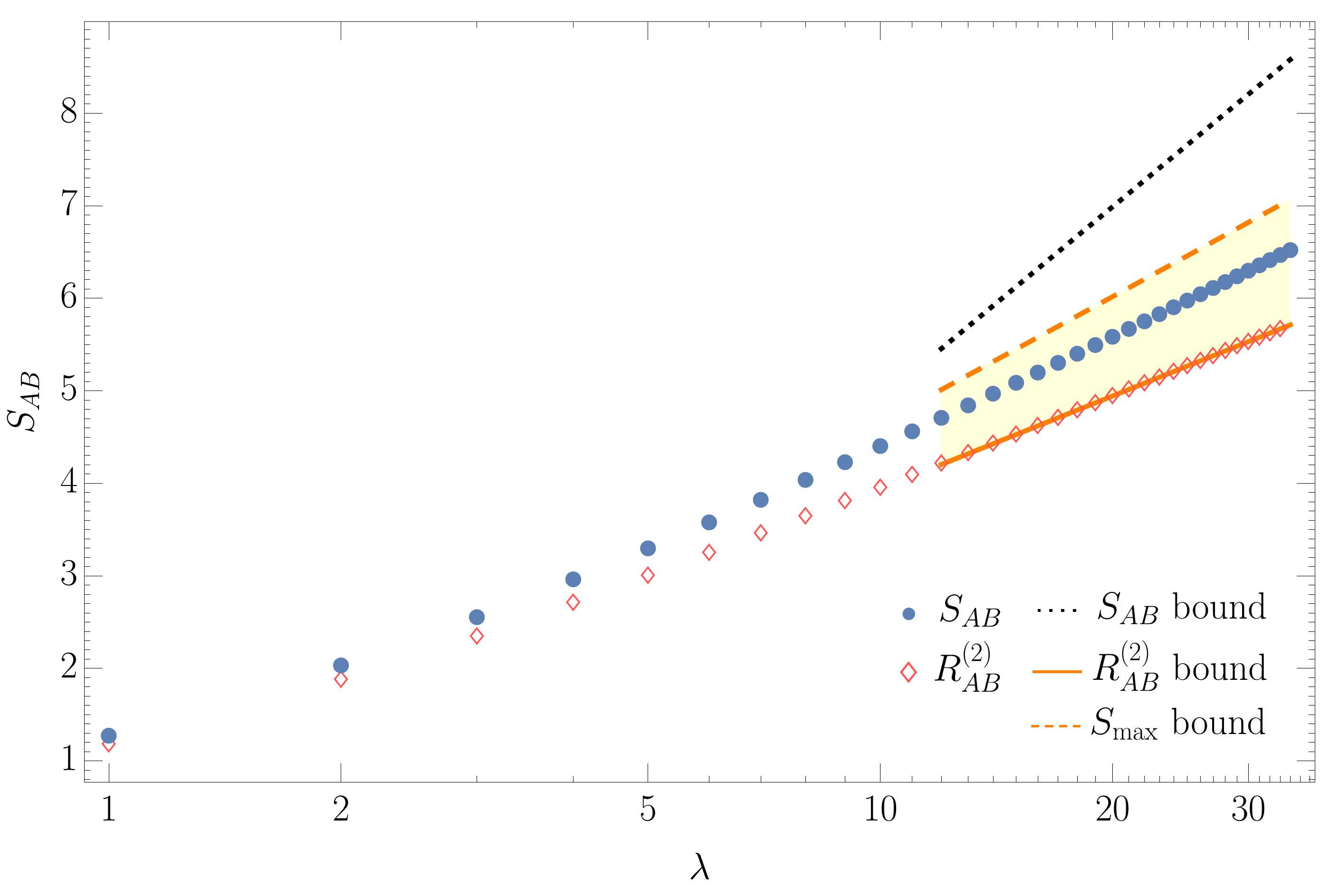}
\caption{[\emph{Graph $\Gamma_6$, entropy of two connected nodes}]. The figure shows our numerical results on the Bell-network state with hexagram graph $\Gamma_6$ and equal spins $j_\ell = \lambda/2$, for a region consisting of two connected nodes. The entanglement entropy $S_{AB}$ is denoted by blue dots, the R\'enyi entropy $R^{(2)}_{AB}$ by red diamonds. The lower bound asymptotic estimate is shown as a solid orange line (the $O(1)$ is fitted using the numerical data). The upper bound estimate given by the maximal entropy is shown as a dashed orange line. We show also the bounds  \eqref{entropy_bounds} as a yellow band. Using the $10$ largest-spin data points, the entanglement entropy $S_{AB}$ is  fitted by $a \log(\lambda) + b + c \lambda^{-1}$ on the last 10 data points obtaining $a\approx 1.85$, $b\approx -0.08$ and $c\approx 2.23$.}\label{fig:hexa2n}
\end{figure}

\paragraph{One node subsystem  of $\Gamma_6$.} Choosing a region $A$ containing a single node, our asymptotic formula reduces to 
\begin{equation}
S_{A} =R^{(p)}_{A}= \log \lambda+ O(1) \, .
\end{equation}
Note that $S_{A}$ and $R^{(p)}_{A}$ can differ by $O(1)$ terms. As already done for the graph $\Gamma_5$, we compute the entanglement entropy and R\'enyi entropy of order two for the hexagram $\Gamma_6$ with all equal spins $j_\ell = \frac{\lambda}{2}$  and scale parameter up to $\lambda\leq 34$. We report the results in Figure \ref{fig:hexa1n}. The plot shows that the $O(1)$ contributions differ for $S_{A}$, $R^{(2)}_{A}$ and the maximal entropy $R^{(0)}_{A} = \log \left(\lambda +1 \right) \approx \log \lambda + O\left(\lambda^{-1} \right)$. We interpret this difference as an indication that the subsystem is not maximally mixed and the Bell-network state is not typical in the Hilbert space. 
\paragraph{Two connected nodes subsystem  of $\Gamma_6$.} Choosing the subsystem $AB$ consisting of two connected nodes, our asymptotic formula reduces to 
\begin{equation}
R^{(2)}_{AB}= \frac{3}{2}\log \lambda+ O(1)\,.
\end{equation}
The asymptotic band with upper and lower bounds on the entanglement entropy given by 
\begin{equation}
\frac{3}{2}\log \lambda \;\leq \; S_{AB} \;\leq\; 3\log \lambda
\end{equation}
For this specific configuration, the maximal entropy $S_{AB} \leq 2\log \lambda$ provides a tighter upper bound. The results of the numerical computation  of the entanglement entropy and R\'enyi entropy of order two for the case of all equal spins $j_\ell = \frac{\lambda}{2}$ and the scale parameter up to $\lambda\leq 34$ are reported in Figure \ref{fig:hexa2n}.

\begin{figure}
\centering
\includegraphics[width=0.9\textwidth]{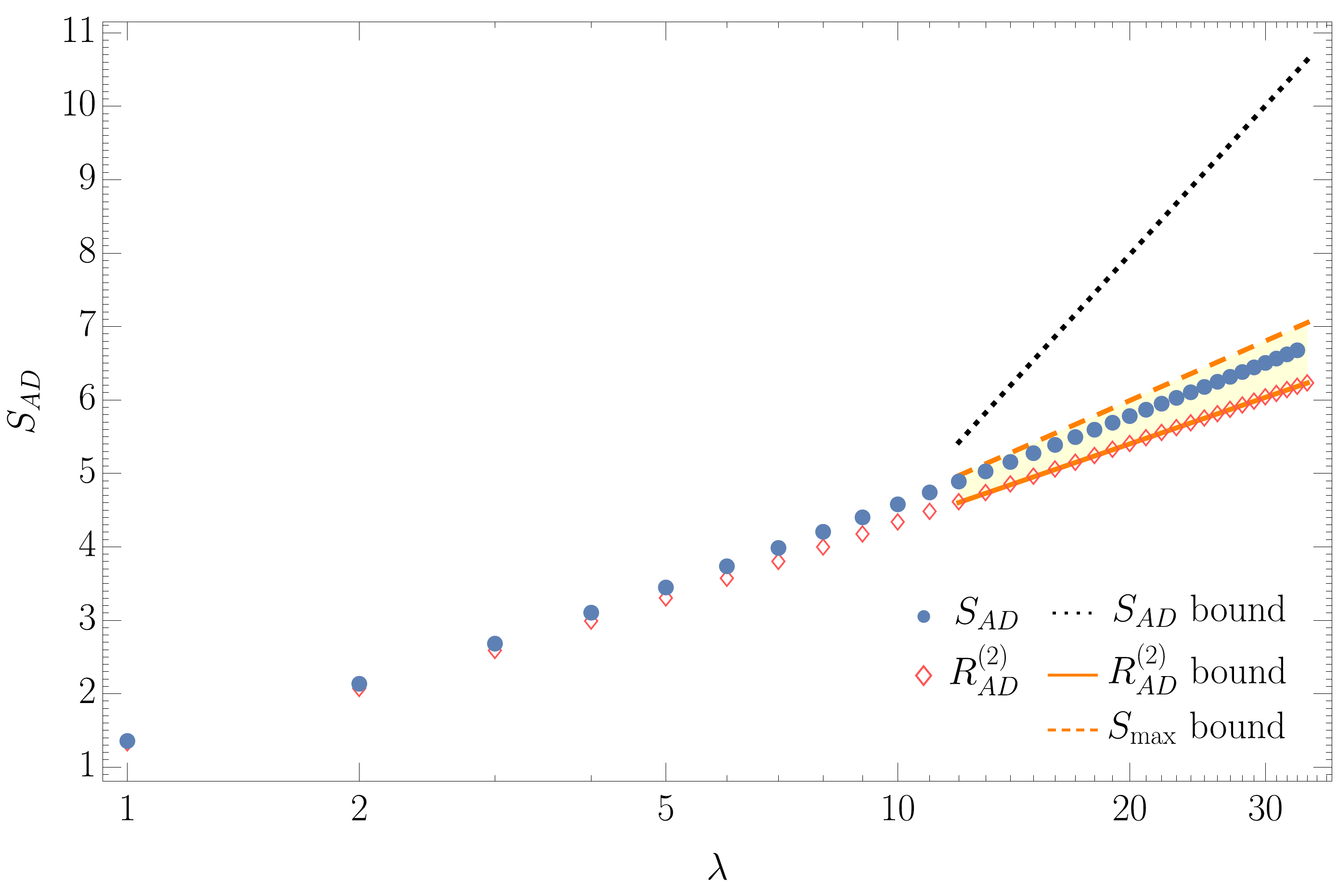}
\caption{[\emph{Graph $\Gamma_6$, entropy of two disconnected nodes}]. The figure shows our numerical results on the Bell-network state with hexagram graph $\Gamma_6$ and equal spins $j_\ell = \lambda/2$, for a region consisting of two \emph{disconnected} nodes. The entanglement entropy $S_{AD}$ is denoted by blue dots, the R\'enyi entropy $R^{(2)}_{AD}$ by red diamonds. The lower bound asymptotic estimate is shown as a solid orange line (the $O(1)$ is fitted using the numerical data). The upper bound estimate given by the maximal entropy is shown as a dashed orange line. We show also the bounds  \eqref{entropy_bounds} as a yellow band. Using the $10$ largest-spin data points, the entanglement entropy $S_{AD}$ is  fitted by $a \log(\lambda) + b + c \,\lambda^{-1}$ on the last 10 data points obtaining$a\approx 1.86$, $b\approx 0.09$ and $c\approx 1.97$.} 
\label{fig:hexa2nb}
\end{figure}

\paragraph{Two disconnected nodes subsystem of $\Gamma_6$.} Choosing a subsystem $AD$ consisting of two disconnected nodes, our asymptotic formula reduces again to 
\begin{equation}
R^{(2)}_{AD}= \frac{3}{2}\log \lambda+ O(1)\,.
\end{equation}
The asymptotic band providing upper and lower bounds for the entanglement entropy is now
\begin{equation}
\frac{3}{2}\log \lambda \;\leq\; S_{AD} \;\leq\; 5\log \lambda\,.
\end{equation}
Again, as the subsystem still consists of a small number of nodes, the maximal entropy $S_{AD} \leq 2\log \lambda$ provides a tighter upper bound. The results of our numerical computations of the entanglement entropy and R\'enyi entropy of order two for the case of all equal spins $j_\ell = \frac{\lambda}{2}$ and the scale parameter up to $\lambda\leq 34$ are reported in Figure \ref{fig:hexa2nb}.

%--------------------------------------------------------------------------------------------------

\section{Discussion}
We studied the entanglement entropy of Bell-network states, numerically and analytically. Bell-network states were introduced in \cite{Baytas:2018wjd} as states that glue quantum polyhedra with entanglement. For given graph $\Gamma$ and spin assignment $j_\ell$, there is a unique Bell-network state $|\Gamma,j_\ell, \mathcal{B}\rangle$ defined by the $SU(2)$ symbol of the graph. Computing their entanglement entropy allows us to put information-theoretic bounds on correlations of shapes of adjacent polyhedra.

\medskip

On the numerical side, we presented a code for evaluating the reduced density matrix of a Bell-network state and its entropy. We use the code to evaluate the entropy for small graphs containing up to six nodes. We consider various subsystems as described in Figure \eqref{fig:graphs}. At fixed graph, we studied spins ranging from $1/2$ to approximately $20$. The numerical results show that Bell-network states are non-typical in the Hilbert space: We find that, at large spins, their entropy remains below by a term $O(1)$ with respect to the one of the maximally mixed state.

While the specific code used here is adapted to small graphs and large spins, a similar procedure can be adopted for general graphs.

\medskip

On the analytical side, we developed methods for computing the R\'enyi entropy of  order $p$ for arbitrary graph and generic region. For spins that are uniformly large, these methods provide reliable bounds on the entanglement entropy.  The R\'enyi entropy is computed using techniques borrowed from spinfoam asymptotics \cite{Barrett:2002ur,Dona:2017dvf}: We write the trace of powers of the reduced density matrix as an integral over unit vectors and $SU(2)$ group elements,
\begin{equation}
\label{summarytrace}
\Tr\left(\rho_A^p\right)= \frac{\int d\vec{n}\, dg \; e^{f_p\left(j,\vec{n},g\right)}}{\left(\int d\vec{n}\, dg \; e^{f_1\left(j,\vec{n},g\right)}\right)^p}\,,
\end{equation}
where $f_p\left(j,\vec{n},g\right)$ is a linear function of all spins $j_\ell$. The integral is then evaluated with saddle-point techniques under a uniform rescaling of the spins $j_\ell \to \lambda j_\ell$ with $\lambda\gg 1$. At the leading order in $\lambda$, we find that the R\'enyi entropy of order $p$ is
\begin{equation}
R_A^{(p)}=\left(\BB -\frac{C^{(p)}_A}{2\left(p-1\right)}\right) \log \lambda + O(1) \,,
\label{eq:Rbound}
\end{equation}
where $\BB$ is the number of links that cross the boundary of the region $A$. The constant $C^{(p)}_A$ is an integer that counts the number of redundant critical-point equations for the `action' $f_p$. While there is no general closed formula, the number $C^{(p)}_A$ can be computed explicitly for a given graph and region, as we have done in the cases that we have studied. Moreover, using our results on the asymptotics of the R\'enyi entropy, we have shown that the entanglement entropy, at leading order in $\lambda$,  scales logarithmically as
\begin{equation}
\left(\BB - \frac{C^{(2)}_A}{2}\right) \log \lambda \, \leq\, S_A \,\leq \, \left(\BB-3\right) \log \lambda \, .
\label{eq:Sbound}
\end{equation}
This result shows that, asymptotically, the entanglement entropy of Bell-network states scales linearly with the number of links $\BB$ that cross the boundary of the region $A$. This result can be understood as an area law for Bell-network states. To clarify this point, let us consider a graph dual to a tessellation of $3$-space and a region $A$. The area of a face dual to a link $\ell$ is $a(j_\ell)=8\pi G\hbar \gamma \sqrt{j_\ell(j_\ell+1)}$. 
Under a rescaling $j_\ell\to\lambda j_\ell$ with $\lambda\gg1$, the area of the boundary of the region $A$ can be written as
\begin{equation}
\textrm{Area}_A=\sum_{\ell\in\partial A}a(\lambda j_\ell)\;= \;\langle a( \lambda j_\ell)\rangle \,|\partial A|,
\end{equation}
where $|\partial A|$ is the number of boundary links and $\langle a( \lambda j_\ell)\rangle$ is the average area of a face. Therefore the entanglement entropy of a Bell-network state takes the form
\begin{equation}
S_A\big(|\Gamma,\lambda j_\ell,\mathcal{B}\rangle\big)\;\approx\; \frac{\log\lambda}{\langle a( \lambda j_\ell)\rangle}\;\textrm{Area}_A\,.
\end{equation}
The origin of this area law is the entanglement between the shapes of quantum polyhedra.\\

We discuss also an application of our numerical results. For the case of a pentagram graph $\Gamma_5$, we find that the entanglement entropy of two adjacent nodes in $|\Gamma_5,\lambda/2, \mathcal{B}\rangle$ is smaller that the sum of the entropies of each node. Calling the two nodes $A$ and $B$, we have $S_A\approx \log \lambda$, $S_B\approx \log \lambda$ and $S_{AB}\approx 1.94\,\log\lambda$, (See Figure \ref{fig:penta2n}). Therefore the mutual information $I(A,B)$ scales as
\begin{equation}
I(A,B)=S_A+S_B-S_{AB}\;\approx\;0.06\,\log\lambda \qquad \qquad (\textrm{Bell-network state})\,.
\label{eq:IAB006}
\end{equation}
This numerical result provides us with a tool to bound correlations of shapes of two adjacent polyhedra in a Bell-network state. Let us consider observables $\mathcal{O}_A$ and $\mathcal{O}_B$ which measure the shape of the quantum polyhedra $A$ and $B$.  In order to have correlated fluctuations of shapes \cite{Bianchi:2006uf,Livine:2006it,Alesci:2008ff,Bianchi:2009ri,Bianchi:2011hp}, the connected correlation function
\begin{equation}
G_{AB}=\langle \mathcal{O}_A\,\mathcal{O}_B\rangle-\langle \mathcal{O}_A\rangle\,\langle\mathcal{O}_B\rangle
\end{equation}
has to be non-vanishing. Remarkably, knowing the mutual information between $A$ and $B$ provides us with a bound on correlations \cite{nielsen00,Wolf},
\begin{equation}
\frac{\big(\langle \mathcal{O}_A\,\mathcal{O}_B\rangle-\langle \mathcal{O}_A\rangle\,\langle\mathcal{O}_B\rangle\big)^2}{2\|\mathcal{O}_A\|^2 \|\mathcal{O}_B\|^2}\leq I(A,B)\,.
\end{equation}
This relation is especially useful for bounded operator with known norm, as is the case for instance for the operator that measures the dihedral angle between two faces of a quantum polyhedron \cite{penrose1971angular,penrose1972nature,Major:1999mc}.

In the case of two quantum tetrahedra in the Bell-network state $|\Gamma_5,\lambda /2, \mathcal{B}\rangle$, our numerical result \eqref{eq:IAB006} tells us that the correlations between shapes are allowed to be non-vanishing at large spins. This result is to be contrasted to the case of the typical state in the Hilbert space $\mathcal{H}_{\Gamma_5\, \lambda/2}$ for which, using Page's result \eqref{eq:Page}, we find
\begin{equation}
I(A,B)=S_A+S_B-S_{AB}\sim\frac{1}{2\lambda} \qquad \qquad (\textrm{Typical state})\,.
\end{equation}
Therefore correlations in shapes of adjacent polyhedra are suppressed as $1/\lambda$ in a typical state, but unsuppressed in a Bell-network state.

The developments presented in this paper are part of an ongoing numerical revolution in the field \cite{Dona:2018nev,Bahr:2016hwc,Bahr:2018gwf,Dittrich:2016tys,Delcamp:2016dqo} and represent the first numerical results on the entanglement entropy of space in loop quantum gravity.

%--------------------------------------------------------------------------------------------------

\section*{Acknowledgments}

E.B. is supported by the NSF Grant PHY-1806428. P.D. is supported by the NSF grants PHY-1505411, PHY-1806356 and the Eberly research funds of Penn State. I.V. thanks Jonathan Engle and U.S. National Science Foundation for partial support under grants PHY-1505490 and PHY-1806290. We thank Simone Speziale for helpful insights at an early stage of this work and Bekir Bayta\c{s} for useful discussions on Bell-network states.

%----------------------------------------------------------------------------
\newpage
\appendix

%--------------------------------------------------------------------------------------------------
\section{Diagrammatic method to compute $f_p\,$: a detailed example}
\label{app:Example}
We report here the detailed construction of $\Tr M^2$ for a pentagram graph and a region $A$ containing two nodes. Using the same graphical notation used in spin foams, each green box corresponds to a $SU(2)$ integral and each closed line corresponds to a $SU(2)$ character: 
\begin{figure}[H]
\centering
$\Tr M^2 \;\;=\;\;$\quad\raisebox{-0.5\height}{\includegraphics[width=0.6\textwidth]{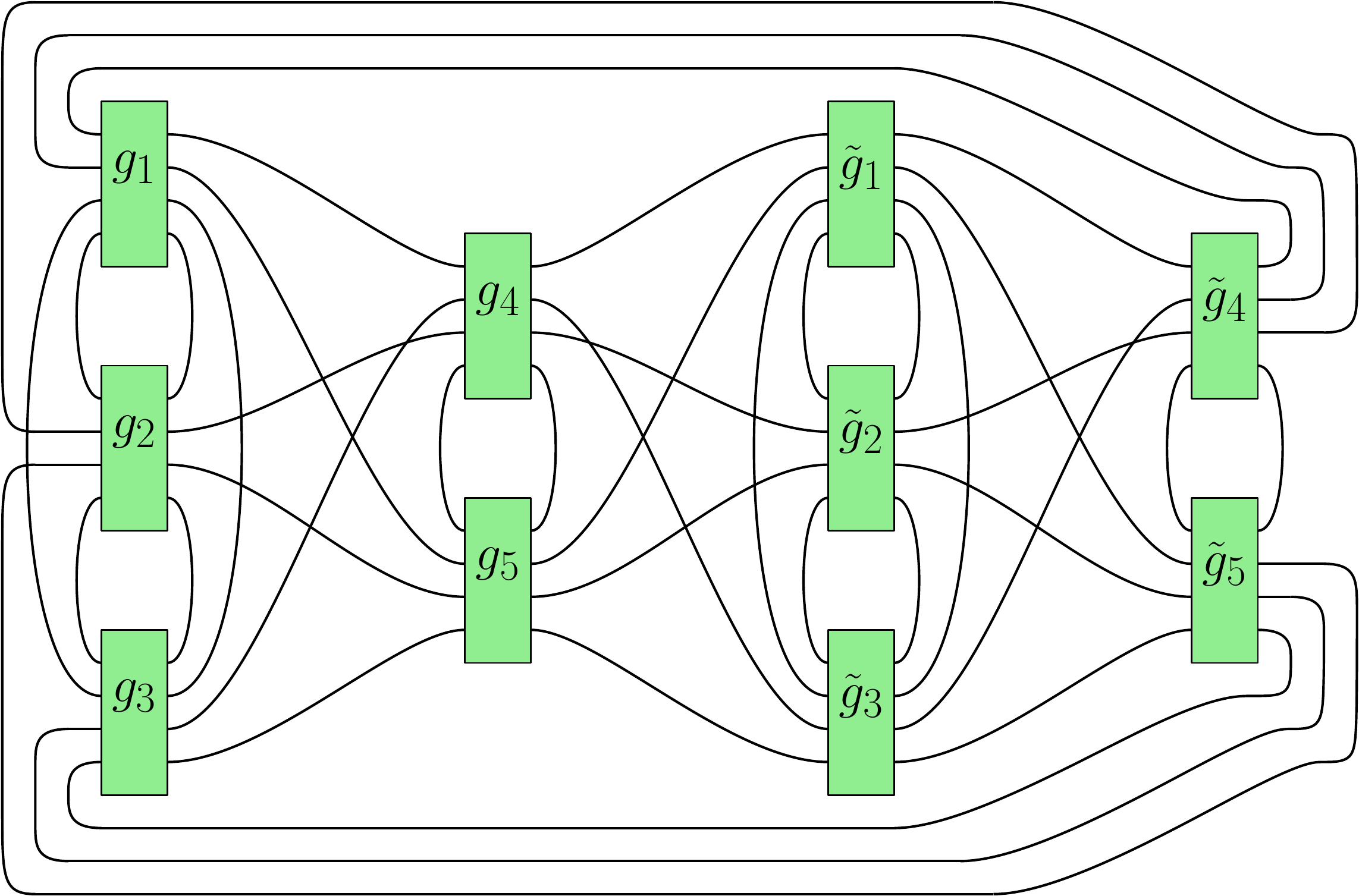}}$\qquad.$
\end{figure}
We call $g_a$ and $\tilde{g}_a$ the group elements,  $j_{ab}$ the spin of the closed line involving the group elements $g_a$ and $g_b$,  $\vec{n}_{ab}$ and $\vec{m}_{ab}$ the unit vectors used to exponentiate the characters as in \eqref{character}. 

The function $f_2$ defined by \eqref{defF}, $\Tr M^2= \int d\vec{n}_f dg_e \,e^{\lambda f_2\left(j_\ell,\vec{n}_f,g_e\right)}$, is given in this case by the expression
\begin{align*}
f_{2}\left(j_{\ell},g_{n},\vec{n}_{f}\right)=&\sum_{a=1}^{3}j_{a4}\,\log\left\langle \vec{n}_{a4}\right|g_{a}g_{4}\tilde{g}_{a}\tilde{g}_{4}\left|\vec{n}_{a4}\right\rangle +\sum_{a=1}^{3}j_{a5}\,\log\left\langle \vec{n}_{a5}\right|g_{a}g_{5}\tilde{g}_{a}\tilde{g}_{5}\left|\vec{n}_{a5}\right\rangle +\\
&\sum_{1\leq a<b\leq3}j_{ab}\,\log\left\langle \vec{n}_{ab}\right|g_{a}g_{b}^{-1}\left|\vec{n}_{ab}\right\rangle +\sum_{1\leq a<b\leq3}j_{ab}\,\log\left\langle \vec{m}_{ab}\right|\tilde{g}_{a}\tilde{g}_{b}^{-1}\left|\vec{m}_{ab}\right\rangle +\\&
j_{45}\,\log\left\langle \vec{n}_{45}\right|g_{4}g_{5}^{-1}\left|\vec{n}_{45}\right\rangle +j_{45}\,\log\left\langle \vec{m}_{45}\right|\tilde{g}_{4}\tilde{g}_{5}^{-1}\left|\vec{m}_{45}\right\rangle 	\, .
\end{align*}

\newpage

\bibliographystyle{JHEPs}
\bibliography{paperEE}

\end{document}